\documentclass[12pt,preprint]{aastex}

\shorttitle{GK star UV band}
\shortauthors{Short and Hauschildt}

\begin{document}


\title{Modeling the near-UV band of GK stars, Paper I: LTE models}


\author{C. Ian Short}
\affil{Department of Astronomy \& Physics and Institute for Computational Astrophysics, Saint Mary's University,
    Halifax, NS, Canada, B3H 3C3}
\email{ishort@ap.smu.ca}

\author{P.H. Hauschildt}
\affil{Hamburger Sternwarte, Gojenbergsweg 112, 21029 Hamburg, Germany}
\email{yeti@hs.uni-hamburg.de}


%\altaffiltext{1}{Visiting Astronomer, Cerro Tololo Inter-American Observatory.
%CTIO is operated by AURA, Inc.\ under contract to the National Science
%\altaffiltext{2}{Society of Fellows, Harvard University.}
%\altaffiltext{3}{present address: Center for Astrophysics,
%    60 Garden Street, Cambridge, MA 02138}
%\altaffiltext{4}{Visiting Programmer, Space Telescope Science Institute}
%\altaffiltext{5}{Patron, Alonso's Bar and Grill}


\begin{abstract}

 We present a grid of LTE atmospheric models and synthetic spectra that 
cover the spectral class range from mid-G to mid-K, and luminosity classes from V
to III, that is dense in $T_{\rm eff}$ sampling ($\Delta T_{\rm eff}=62.5$ K),
for stars of solar metallicity and moderately metal poor scaled solar 
abundance ($[{{\rm A}\over{\rm H}}]=0.0$ and $-0.5$).  All models have
been computed with two choices of atomic line list: a) the ``big'' line lists
of \citet{kurucz92a} that best reproduce the broad-band solar blue and 
near UV $f_\lambda$ level, and b) the ``small'' lists of \citet{kuruczp75} 
that provide the best fit to the high resolution solar blue and near-UV
spectrum.  We compare our model SEDs to a sample of stars carefully selected
from the large catalog of uniformly re-calibrated spectrophotometry of 
\citet{burn} with the goal of determining how the quality of fit  
varies with stellar parameters, especially in the historically troublesome 
blue and near-UV bands.  We confirm that our models computed with the ``big'' 
line
list recover the derived $T_{\rm eff}$ values of the PHOENIX NextGen grid, 
but find that the models computed with the ``small'' line list provide
greater internal self-consistency among different spectral bands, and 
closer agreement with the empirical $T_{\rm eff}$ scale of \citet{ramirezm05},
but not to the interferometrically derived $T_{\rm eff}$ values of 
\citet{baines10}.  We find {\it no}
evidence that the near UV band discrepancy between models and observations
for Arcturus ($\alpha$ Boo) reported by \citet{shorth03} and \citet{shorth09} 
is pervasive, and that Arcturus may be peculiar in this regard.

\end{abstract}



\section{Introduction}


%% Included in this acknowledgments section are examples of the
%% AASTeX hypertext markup commands. Use \url without the optional [HREF]
%% argument when you want to print the url directly in the text. Otherwise,
%% use either \url or \anchor, with the HREF as the first argument and the
%% text to be printed in the second.

Critical comparisons of the observed absolute spectral energy distribution 
(SED), $f_\lambda(\lambda)$, to that predicted with PHOENIX computational
models from the near UV to the red have been carried out for the Sun, Procyon,
and Arcturus (see \citet{shorth09}, \citet{shorth05}, and \citet{shorth03}).
They presented evidence that both LTE and non-LTE models increasingly
over-predict the $f_\lambda$ level for $\lambda < 4000$ \AA~ as the 
$T_{\rm eff}$ value of the star decreases from that of early G stars
to that of early K.  The investigations to date have been restricted to
bright standard stars for which there is very high quality spectroscopic
and spectrophotometric data, with the consequence that the quality 
of the SED fit has not been well sampled in the stellar parameter space
{$T_{\rm eff}/\log g/{\rm [}{{\rm A}\over{\rm H}}{\rm ]}$}.  Here, we take the first
major step to rectify the 
situation by comparing a large grid of model SEDs spanning the cool side of the 
HR diagram to observed SEDs taken from the spectrophotometric catalog of
\citet{burn}.  Our goal is to map out the goodness of fit, and
the magnitude of any systematic discrepancies between model and observed SEDs as a
function of the three stellar parameters, $T_{\rm eff}$, $\log g$, and
[${{\rm A}\over{\rm H}}$], with the more fundamental goal of constraining
the physical character of the sources of any discrepancies.  We also compare
our $T_{\rm eff}$ values inferred from SED fitting to other empirical
and theoretical $T_{\rm eff}$ calibrations.

\section{Observed $f_{\lambda}(\lambda)$ distributions \label{sobsseds}}

\citet{burn} presented a large catalog (henceforth B85) of observed SEDs taken with 
photo-electric instruments on 0.5m class telescopes at 
various observatories in the former USSR from the late 1960s to the mid 
1980s and uniformly 
photometrically re-calibrated to the ``Chilean system'' (\citet{shorth09} 
contains a more detailed description of the individual data sources
included in this compilation).    
These data sets
all generally cover the $\lambda$ range 3200 to 8000 \AA~ with 
$\Delta\lambda=25$ \AA, and have a quoted ``internal
photometric accuracy'' of $\approx3.5\%$.  

\paragraph{}

 We have harvested from the B85 catalog a sample of stars that meet the
following criteria: 1) having an entry in the
5$^{\rm th}$ Revised Edition of the Bright Star Catalogue \citep{hoffleitw91}, 
(henceforth BSC5), 2) having a spectral class later than G0 and earlier 
than M0,
3) having a luminosity class in the range from V to III, 4) having
an entry in the metallicity catalog of \citet{cayrelsr01}, 5) having
no variability, chemical peculiarity, or binarity flags in BSC5, and 6) 
having an $f_\lambda$ distribution that was not obviously
inconsistent based on visual inspection, with those of other stars of the
same BSC5 spectral class and within $\pm 1$ sub-class.  The spectral and luminosity
classes were taken from The Revised Catalog of MK Spectra Types for the Cooler Stars
\citep{keenan_n00}, the paper of \citet{keenan_b99}, or The Perkins Catalog of Revised MK Types for 
the Cooler Stars \citet{keenan_m89}, in decreasing order of preference.  For
nine objects without spectral classes determined by Keenan or his collaborators,
we took spectral types from the General Catalogue of Stellar Spectral Classifications 
(Version 2010-Mar) \citet{skiff10}: HD 50522 \citep{abt08}, HD 147675 \citep{torres06}, \citep{houk75}, HD 71878 \citep{gray06},
HD 222107 \citep{gray03}, HD 221673 \citep{abt81}, HD 156266, 115383 \citep{harlan74}, HD 34559, 49878 \citep{adams35}.  
For the six objects for which \citet{skiff10}
had multiple conflicting spectral classes, we took the class that agreed with that
listed in BSC5.  Application of criterion 6) 
necessitated only including stars for which there was more than one object of
the same spectral class, to within $\pm 0.5$ sub-classes.
We found that application of these six criteria, along with the metallicity 
criteria described below, limited us to 33 stars of spectral class
ranging from G5 to K4 for giants and G0 to G5 for dwarfs.  All spectra were
corrected for their heliocentric radial velocity, {\it RV}, using the RV values in
BSC5.  However, we expect the RV correction to have a very minor effect on the quality 
of spectral fitting at the low spectral resolution of the B85 data.  

\paragraph{}

 We restricted the sample to two metallicity ranges
based on the value of [${{\rm A}\over{\rm H}}$] reported in the catalog
of \citet{cayrelsr01}: a) stars of $-0.1 < [{{\rm A}\over{\rm H}}] < 0.1$,
and b) stars of $-0.4 < [{{\rm A}\over{\rm H}}] < -0.6$.  These two 
ranges are expected to contain stars of measured [${{\rm A}\over{\rm H}}$]  
approximately equal to 0.0 and -0.5, respectively, on the grounds that
the quoted uncertainties, $\Delta [{{\rm A}\over{\rm H}}]$, of 
determinations of {\it overall} metallicity typically have quoted uncertainties of the order 
of $\pm 0.1$.  For stars 
that contained multiple entries in the catalog of \citet{cayrelsr01}, we
based the metallicity selection on either the mean, or the median, value of 
[${{\rm A}\over{\rm H}}$], depending on whether there were obvious outlier 
values and skew.  Metallicity range b) approximately represents a moderately metal 
poor population typical of the older thick disk, represented by stars such as Arcturus 
([${{\rm A}\over{\rm H}}$]$\approx -0.7$).  Stars of 
[${{\rm A}\over{\rm H}}$]$ < -0.5$ are increasingly rare in B85 catalog 
as [${{\rm A}\over{\rm H}}$] decreases, so -0.5 was the lowest 
[${{\rm A}\over{\rm H}}$] value for which we could harvest a significant
number of stars of [${{\rm A}\over{\rm H}}$] value within $\pm 0.1$ of
each other.
Applying the six criteria described above, we harvested 30 and
3 stars from the B85 catalog in each of the metallicity ranges a) and b), 
respectively.  Table \ref{tabb85} contains the list of the stars that 
were selected in each metallicity range, and the number of 
[${{\rm A}\over{\rm H}}$] measurements in the \citet{cayrelsr01} catalog. 
Some stars had two independently
measured SEDs in the B85 catalog, and where that is the case, both
spectra are included in our analysis and provide a check on the internal consistency
of the B85 spectrophotometry.  The number of spectra is also indicated
in Table \ref{tabb85}.  

\paragraph{}

\begin{figure}
%\plotone{/disc/disc51/ishort/Paper5/G5IIIStars.eps}
\plotone{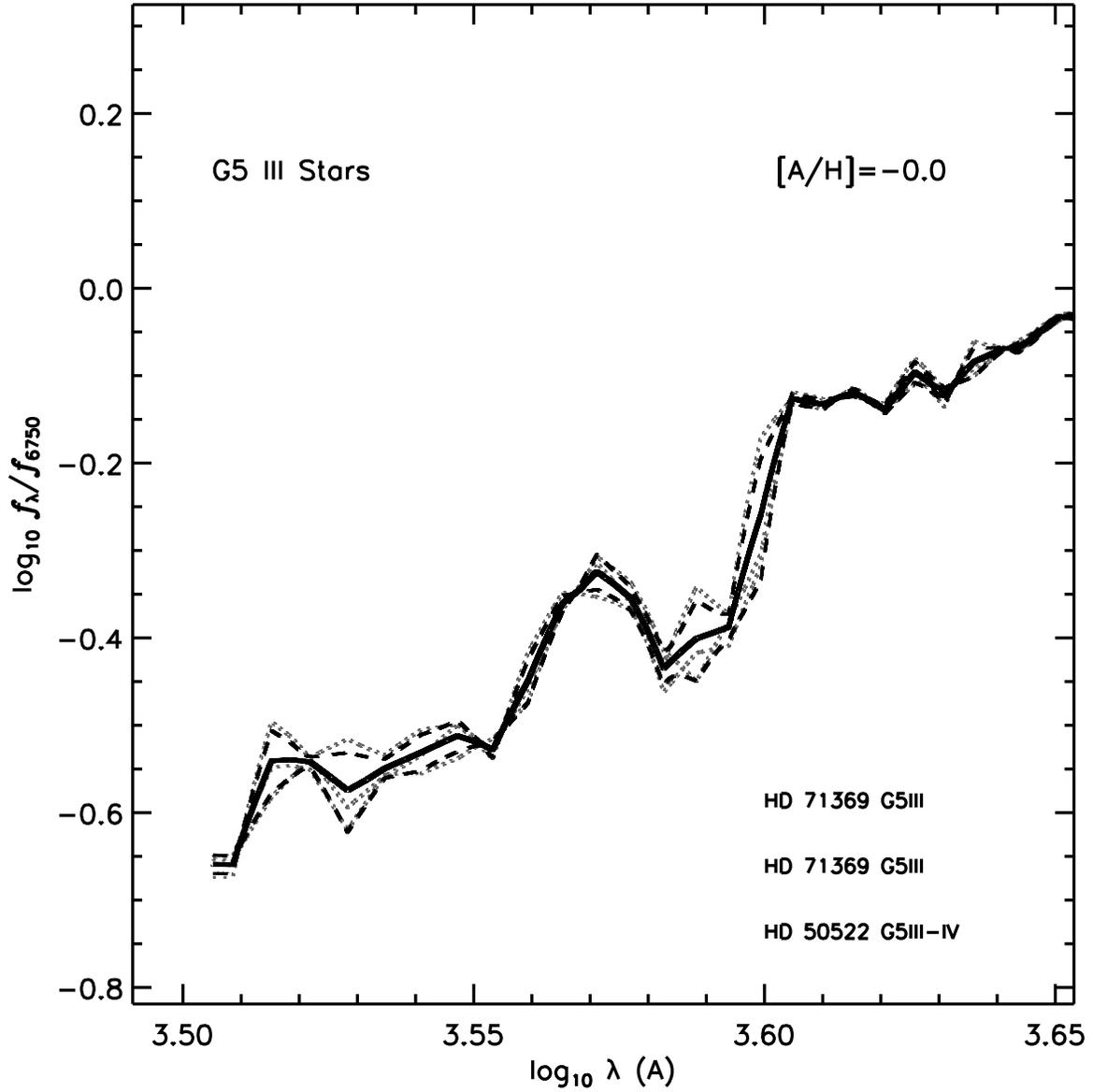}
\caption{G5 III sample (two stars, three spectra).  Light gray dotted lines: individual 
normalized stellar spectra, $f_{\lambda, {\rm 6750}}$ (see text), from 
catalog of B85; black solid line: sample average $f_{\lambda, {\rm 6750}}$ spectrum; black dashed lines: 
$\pm 1~ \sigma$ $f_{\lambda, {\rm 6750}}$ spectra.  \label{fvaryG5III}}
\end{figure}

\begin{figure}
%\plotone{K3-4IIIStars.eps}
\plotone{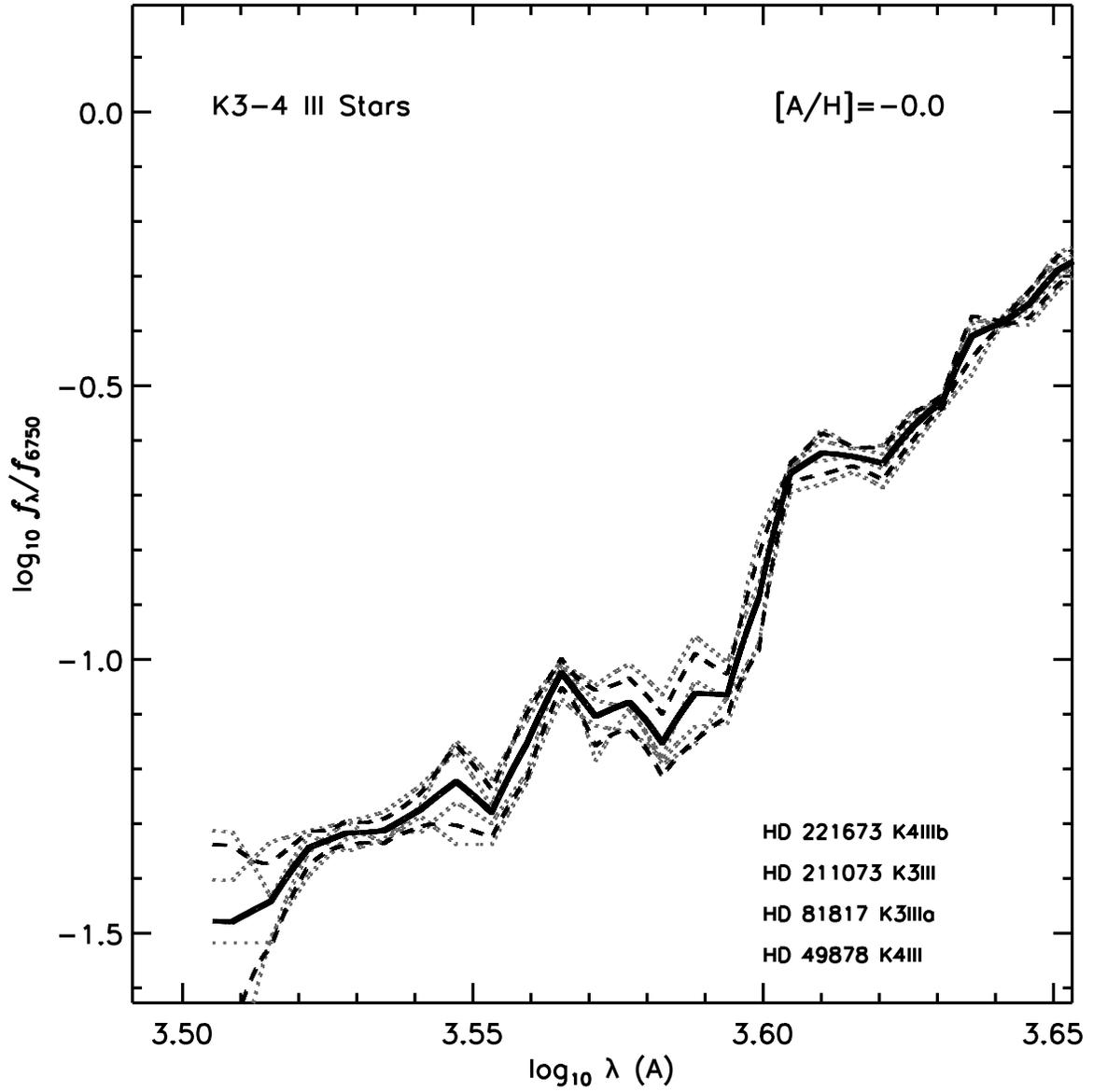}
\caption{Same as Fig. \ref{fvaryG5III}, but for the K3-4 III sample (four stars, four spectra).  
  \label{fvaryK4III}}
\end{figure}

Figs. \ref{fvaryG5III} and \ref{fvaryK4III} show, for the hottest and coolest giant samples, the 
individual spectra of all stars
of that sample, over-plotted with the sample mean spectrum and the spectra of
 the $\pm 1 \sigma$ deviation from the mean.  Note that the $\sigma$ values are most
clearly meaningful for the two samples that contain more than a few spectra, namely,
the G8 III and K0 III stars of solar metallicity.  For most samples, the spectra that
passed our selection procedure fall within $\pm 1 \sigma$ of the mean at
most $\lambda$ values.  
We have separately calculated mean and deviation spectra for each sub-sample. 
%unnecessary?? and show the results in Figs. \ref{fvaryK0IIIl} and \ref{fvaryK0IIIh}.  

\subsection{K0 III sample}

\begin{figure}
%\plotone{K0IIIStars.eps}
\plotone{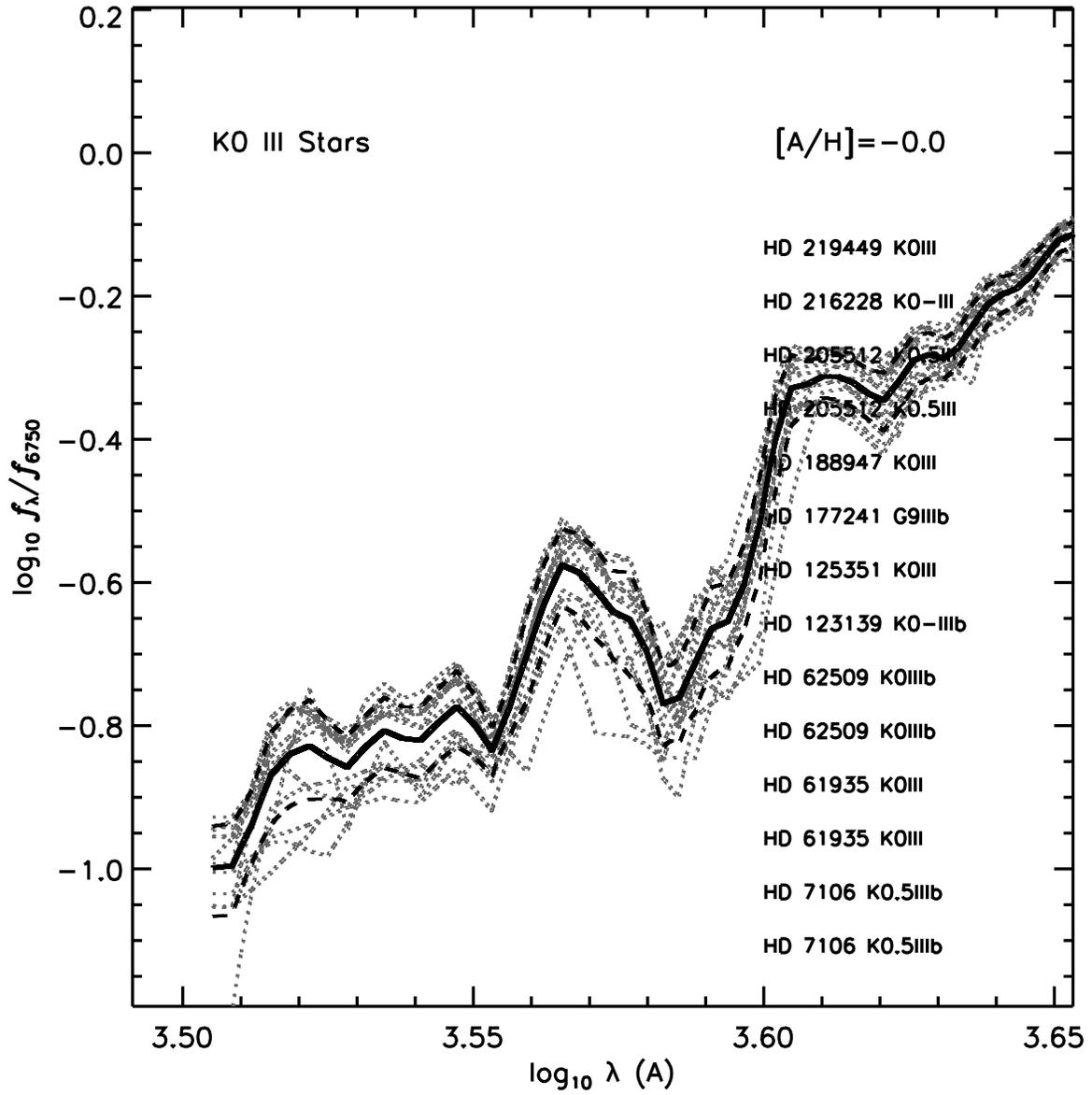}
\caption{Same as Fig. \ref{fvaryG5III}, but for the K0 III sample (ten stars, fourteen spectra).  The 
spectra fall into two groups in the $3.50 < \log\lambda < 3.58$ range, those of ``high'' and those
of ``low'' UV flux, with an apparent dearth of stars of intermediate UV flux level.
  \label{fvaryK0III}}
\end{figure}

The K0 III sample, shown in Fig. \ref{fvaryK0III} is special in that it 
contains more spectra
than the other sub-classes (14 spectra of ten objects), and the distribution of spectra 
has a bifurcation
that can be most clearly seen in the $\log\lambda$ region from 3.52 to 3.55 (3300 to 3550 \AA).
  Therefore, we have broken up the K0 III sample into two sub-samples, an $l$ sub-sample 
 of four stars
(six spectra) with ``low'', and an $h$ sub-sample with six stars (eight spectra) with 
``high'', UV $f_\lambda$ level, respectively.  The
stars belonging to the $l$ and $h$ sub-samples are indicated in Table \ref{tabb85}. 
This bifurcation may be an artifact of our sorting of stars with an effective precision of 
about one spectral sub-class.  Around K0, stars can be typed to half-sub-class precision
(see, for example, \citet{keenan_n00}), and differences of $\approx 0.5$ spectral 
sub-classes would presumably have the greatest effect in the blue and near UV spectral 
bands.  We note that two of the four stars in our ``low'' sample have a spectral class of 
K0.5 III.  

\subsection{Arcturus}

To compare with the results of \citet{shorth03}, it is necessary to include 
Arcturus ($\alpha$ Boo. HD124897) (and other stars of the same SED, were they 
available), in the current investigation. 
The metallicity catalog of \citet{cayrelsr01} lists 17 measurements of 
[${{\rm A}\over{\rm H}}$] for Arcturus
ranging from -0.370 to -0.810, with most values in the range of -0.4 to -0.6, and the average
being -0.54.
Arcturus is the {\it only} star in the B85 catalog with a spectral class in the range 
from K0 to K2 III with 
measured $[{{\rm A}\over{\rm H}}]$ values within $\pm 0.2$ dex of -0.5.  Therefore,
we cannot form a K1 III sample at $[{{\rm A}\over{\rm H}}]\approx -0.5$ for quality
control inspection, or for forming a mean spectrum, as we have with the other
spectral classes and metallicities.   
Moreover, Arcturus is flagged in BSC5 as being chemically peculiar, with the spectral 
class designation being K1.5IIIFe-0.5.   

 \paragraph{}
 With an apparent $V$ band magnitude of -0.04, the B85 spectrophotometric data for Arcturus 
should be of relatively high quality.  
The B85 catalog contains three measured $f_\lambda$ distributions, two 
of which are consistent with each other on the basis of visual inspection.
The third is higher by about 0.05 dex for $\log\lambda$ less than about 3.55 
($\lambda<3550$ \AA).  We have formed a sample at spectral class K1.5 III for
stars of nominal $[{{\rm A}\over{\rm H}}]$ equal to -0.5 using these three SEDs.

\section{Model grid}

\subsection{Atmospheric structure calculations}

 We have computed a grid of about 400 atmospheric models under 
the approximation of LTE with spherical $1D$ geometry spanning a range
in $T_{\rm eff}$ from 6250 to 4000 K with a sampling, $\Delta T_{\rm eff}$, of 125 K, 
and in $\log g$ from 1.5 to 5.0 with a sampling of 0.5, for [${{\rm A}\over{\rm H}}$] values
of 0.0 and -0.5.  A $\Delta T_{\rm eff}$ interval of 125 K is close to the nominal 
$T_{\rm eff}$ difference between successive spectral sub-classes among GK stars (see, for
example, \citet{ramirezm05}).  Furthermore, we interpolate in our synthetic spectrum grid
to achieve an effective $\Delta T_{\rm eff}$ of 62.5 K, as described below. 
%Based on preliminary experience comparing an initial, coarser grid to the
%observed SEDs harvested from the B85 catalog, the sampling of our finer grid varies,
%becoming sparser in regions of parameter space unoccupied by observed objects.  

\paragraph{}

For spherical models,
a value of the effective radius, $R_{\rm eff}$, at $\tau_{\rm 12000}=1$ is also required as
input.  However, we expect the value of $R_{\rm eff}$ will have a minor impact on
the computed SED compared to $T_{\rm eff}$, $\log g$, and [${{\rm A}\over{\rm H}}$], and we used values 
corresponding to a mass of $1 M_{\rm Sun}$ at the given value of $\log g$ for all models.  Because
we are restricted to stars in the spectral class range G5 to K5, most of which are of luminosity
class III (see Section \ref{sobsseds}), we expect the mass distribution of the progenitor main
sequence objects to be of the order of $1 M_{\rm Sun}$.  \citet{hauschildtafba99} 
found
that for supergiant models of given $T_{\rm eff}$ and $\log g$, which are even
more spherically extended than our models. the choice of mass in the
range 2.5 to 7.5 $M_{\rm Sun}$ lead to 
only small differences in the {\it relative} SED.  

\paragraph{}

\begin{figure}
%\plotone{/disc/disc51/ishort/Paper5/xi_comp.eps}
\plotone{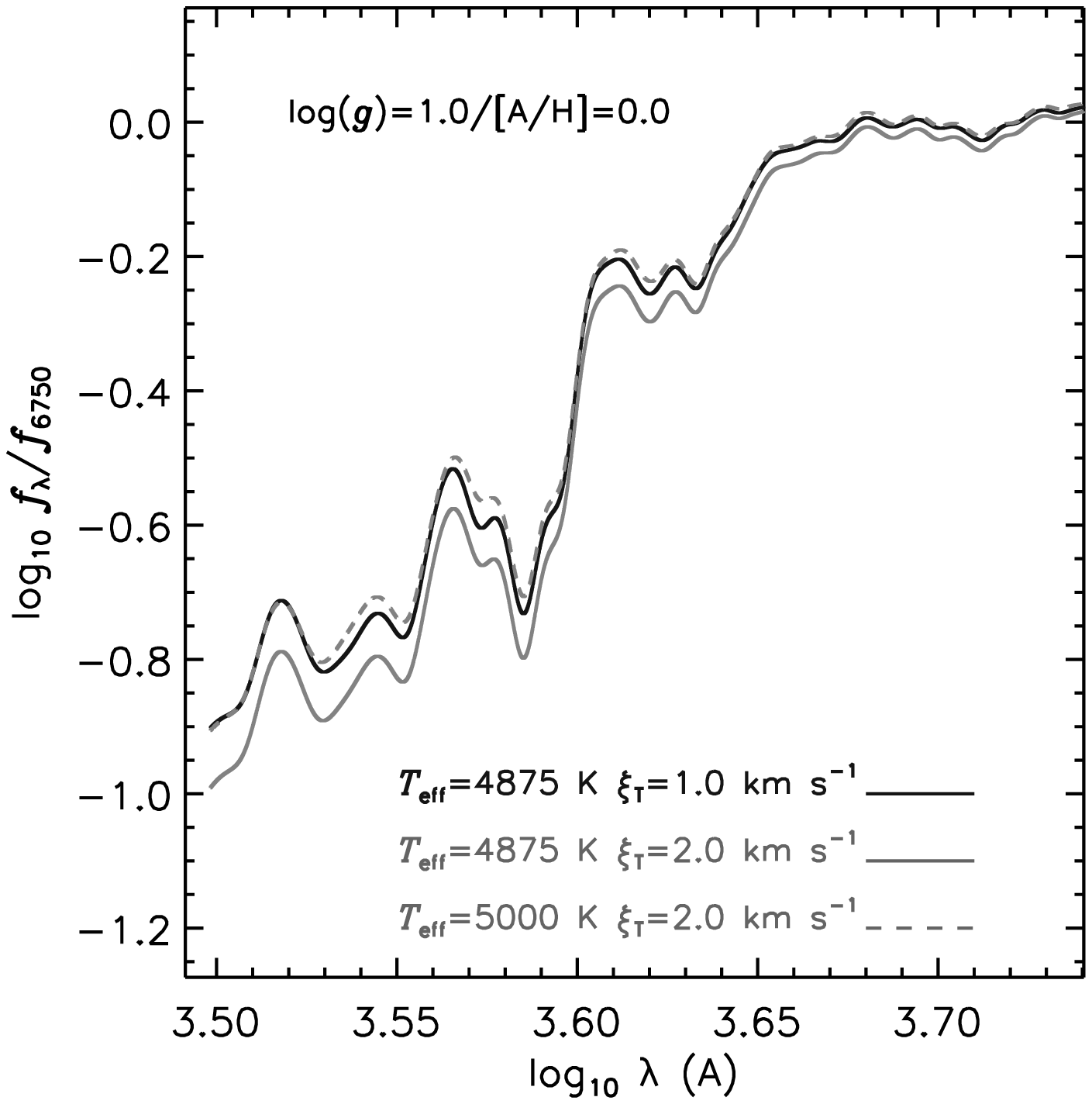}
\caption{Synthetic SEDs for models of varying $T_{\rm eff}$ and $\xi_{\rm T}$
for $\log g=2.0$ and $[{{\rm A}\over{\rm H}}]=0.0$.
Solid dark line: model of $T_{\rm eff}=4875$ K and $\xi_{\rm T}=1.0$ km s$^{\rm -1}$,  
solid light line: model of $T_{\rm eff}=4875$ K and $\xi_{\rm T}=2.0$ km s$^{\rm -1}$, 
dashed line: model of $T_{\rm eff}=5000$ K and $\xi_{\rm T}=2.0$ km s$^{\rm -1}$.
\label{microt} }
\end{figure}

The observed SED will depend on the
value adopted for the micro-turbulent velocity dispersion, $\xi_{\rm T}$, which determines the extent
to which spectral lines are blended in crowded spectral regions, such as the blue band, and therefore, 
affects the $f_\lambda$ level to different extents in different $\lambda$ regions.  We adopted 
values of 1.0 and 2.0 km s$^{-1}$ for $\log g$ values greater than and less than 3.5,
respectively.  \citet{shorth05} found that $\xi_{\rm T}=1$ km s$^{-1}$ provided a good fit
to the solar SED from the near UV to the red.  From analysis of line profiles in high resolution high 
$S/N$ spectra of G and K III stars, previous investigators have found $\xi_{\rm T}$ ranging from 1.5 to 
2.0 km s$^{-1}$ \citep{gray82}, \citep{foy78}, \citep{gustafsson_ka74}.  In Fig. \ref{microt} we 
compare two synthetic SEDs computed with $\xi_{\rm T}=1$ and $2$ km s$^{-1}$ for a model of $T_{\rm eff}=4875$ K,
$\log g=2.0$, and [${{\rm A}\over{\rm H}}$]=0.0, and a model of $\xi_{\rm T}=2$ km s$^{-1}$ and 
the same stellar parameters except that $T_{\rm eff}$ is 5000 K.  The value
of $\xi_{\rm T}$ makes an increasingly large difference as $\lambda$ decreases because
of the increasing density of spectral absorption lines, and reaches a maximum of $\approx 0.1$
dex at 3200 \AA.  This is a significant effect, and a proper exploration of the effect on the 
stellar parameters
determined from SED fitting will require a model grid with a $\xi_{\rm T}$ dimension, and is
beyond the scope of the current investigation.  However, we can note qualitatively from Fig. \ref{microt}
that models 
models in which $\xi_{\rm T}$ is over-estimated by 1 km s$^{-1}$ will produce fits
to the blue and near UV band flux level that over-estimate the value of $T_{\rm eff}$
by $\approx 100$ K.  For all models, we adopted
a mixing length parameter for the treatment of convective flux transport of one pressure
scale height.

\paragraph{}

For all models, we adopt scaled solar abundances.  There is evidence that mildly metal poor
stars of [${{\rm A}\over{\rm H}}$]$=-0.5$ have abundance distributions that are enhanced in 
$\alpha$-elements with respect to the Sun by 0.2 to 0.3 dex (see, for example, \citet{pdk93}).  However, given the scope
of the model grid required for this initial investigation, we have decided to restrict ourselves to
scaled solar ${{\rm A}\over{\rm H}}$ distributions.  There has been recent controversy over the values of
the solar abundances for important elements such as CNO, with \citet{asplund04} 
and other papers in that Series, revising the values
downward by 0.2 to 0.3 dex on the basis of $3D$ models that account for hydrodynamic turbulence.  However, the work of 
\citet{asplund04} has been found to be discrepant with abundances determined from very sensitive helioseismological 
fitting, and has recently been thrown into question on the grounds of self-consistency (see, for example, 
the very thorough recent investigation by \citet{pinsonneault09}).  For 
simplicity, we restrict ourselves in this investigation to the solar 
abundance distribution of \citet{grev_ns92}.    
 
\paragraph{}

For each grid point, we compute two models that differ from each other in the choice of atomic
line list.  Our Series 1 and 2 models use the ``big'' and ``small'' line lists 
of \citet{shorth09}, respectively, and that paper contains a description of the content of the two line lists, the
evidence for and against each one, and an investigation of the effect that the choice of line list 
has on the computed SED, especially in the heavily line blanketed blue and near UV region.  \citet{shorth09}
found that the choice of ``big'' or ``small'' line list has a significant impact on the computed
$f_\lambda$ values in the blue and near UV bands for both the Sun and Arcturus, and concluded that NLTE 
models of the Sun with the ``big'' line list provide a better fit to the measured $f_\lambda$ distribution 
of \citet{necklabs}, but that those with the ``small'' line list provide a better  
fit to three other measured $f_\lambda$ distributions, including one measured from space.  Therefore, 
we have decided to evaluate the fit provided by models with both choices of line list here.  We note 
that the choice of line list affects the model structure as well as directly affecting the synthetic 
SED calculation.

\subsection{Synthetic spectra}

For each model in both Series 1 and 2 we computed self-consistent synthetic spectra in the
$\lambda$ range 3000 to 8000 \AA~ with a spectral resolution 
($R={\lambda\over\Delta\lambda}\approx 350\,000$)
to ensure that spectral lines were adequately sampled.
These were then degraded to match the low resolution measured $f_\lambda$ distributions of B85
by convolution with a Gaussian kernel.  
B85 present their data with a sampling of 25 \AA, but, neither B85, nor any of the original source
publications that are still accessible, describe the instrumental spectral profile.  Based on trial 
comparisons of observed SEDs with synthetic $f_\lambda$ spectra convolved with Gaussians
of various FWHM values, we found that a Gaussian instrumental profile with a FWHM value of 75 \AA~ 
provided the closest qualitative match to the spectral structure in the observed SEDs.  Therefore,
we have convolved all our synthetic spectra with a normalized Gaussian kernel of FWHM equal to
75 \AA.  We note that this convolution also automatically accounts for macro-turbulence, which 
has been found to be around 5.0 km s$^{\rm -1}$ for G and K II stars \citet{gray82}.  
%%% PROBABLY NOT WORTH MENTIONING The wavelength values of the synthetic SEDs were also corrected for the refractive index of air.  
We interpolate in $\log T_{\rm eff}$ between pairs of synthetic SEDs to obtain a grid
with an effective sampling, $\Delta T_{\rm eff}$, of 62.5 K.  A $T_{\rm eff}$ difference of
62.5 K at 3300 \AA~ corresponds to a difference in the $\log f_\lambda$ of a blackbody of 0.07,
and this difference will be smaller for larger $\lambda$ and higher $T_{\rm eff}$ values.
For comparison, the discrepancy in near UV band $\log f_\lambda$ level between
LTE models and observations found by \citet{shorth09} for Arcturus around 3300 \AA~ 
is $\approx 0.15$. 

\paragraph{}

\begin{figure}
%\plotone{G5IIIStars.uv.eps}
\plotone{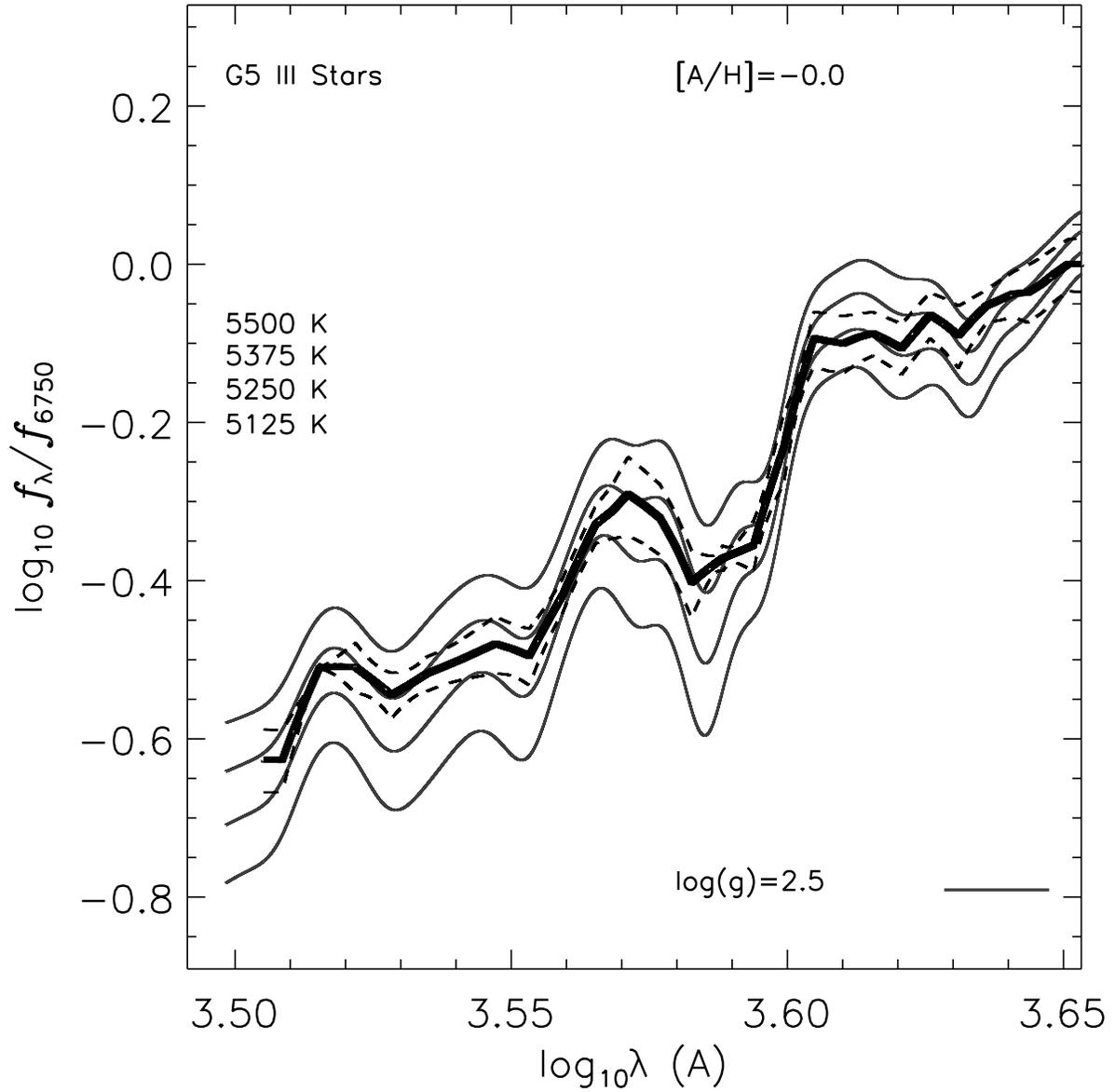}
\caption{G5 III sample: Comparison of sample average to synthetic $f_{\lambda, {\rm 6750}}$ spectra
of Series 1 models.  
Solid black line: sample average $f_{\lambda, {\rm 6750}}$ spectrum, black dashed lines: 
$\pm 1~ \sigma$ spectra, solid gray lines: closest matching and bracketing synthetic $f_{\lambda, {\rm 6750}}$ spectra. 
  \label{fcompG5III}}
\end{figure}

\begin{figure}
%\plotone{K3-4IIIStars.uv.eps}
\plotone{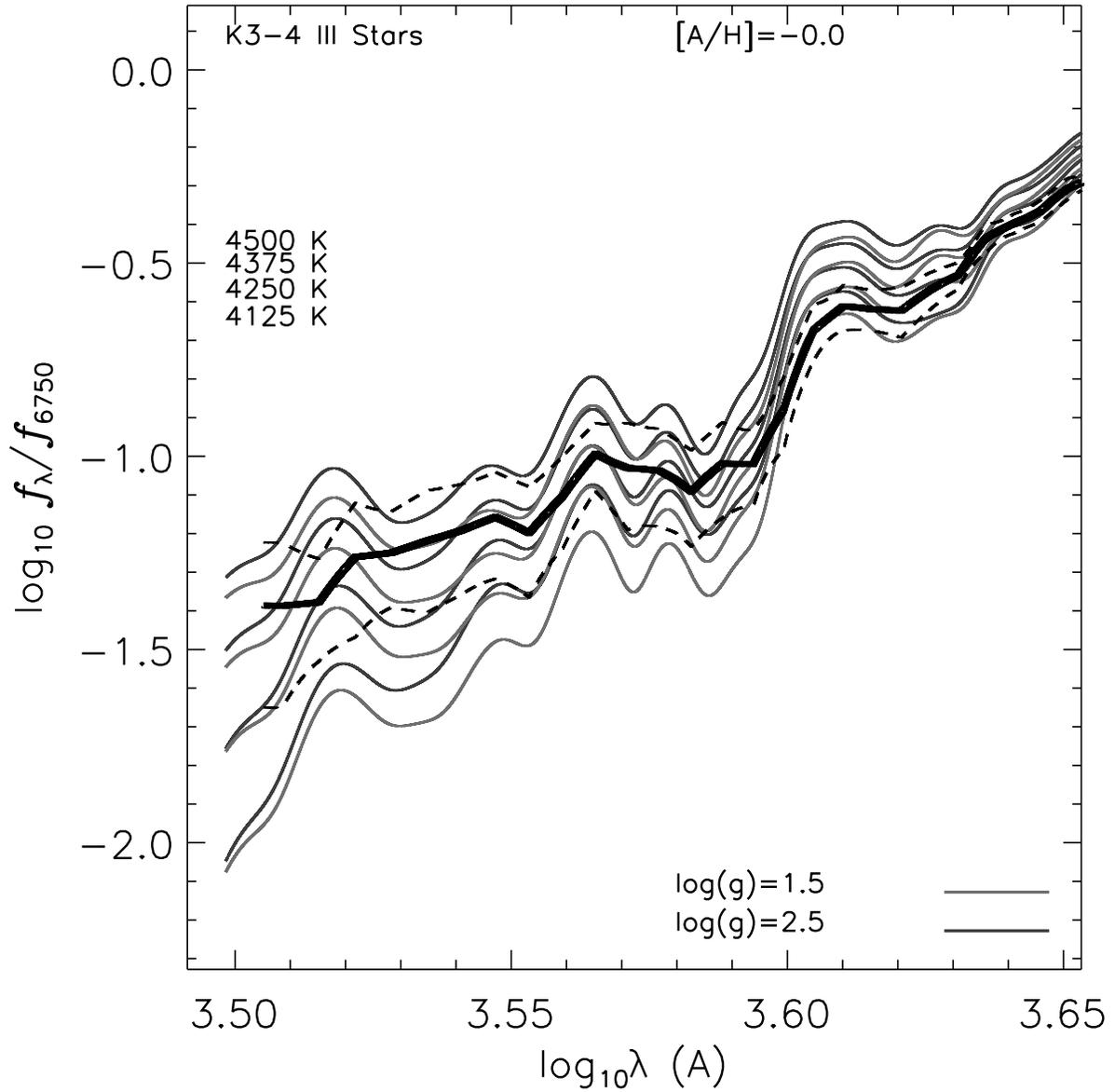}
\caption{Same as Fig. \ref{fcompG5III}, but for the K3-4 sample.  Medium gray lines: $\log g=2.5$, light gray
lines: $\log g=1.5$.  For clarity, the synthetic spectra of the $\log g=2.0$ models have been omitted.
 \label{fcompK4III}}
\end{figure}

\begin{figure}
%\plotone{G5IIIStars.diff.eps}
\plotone{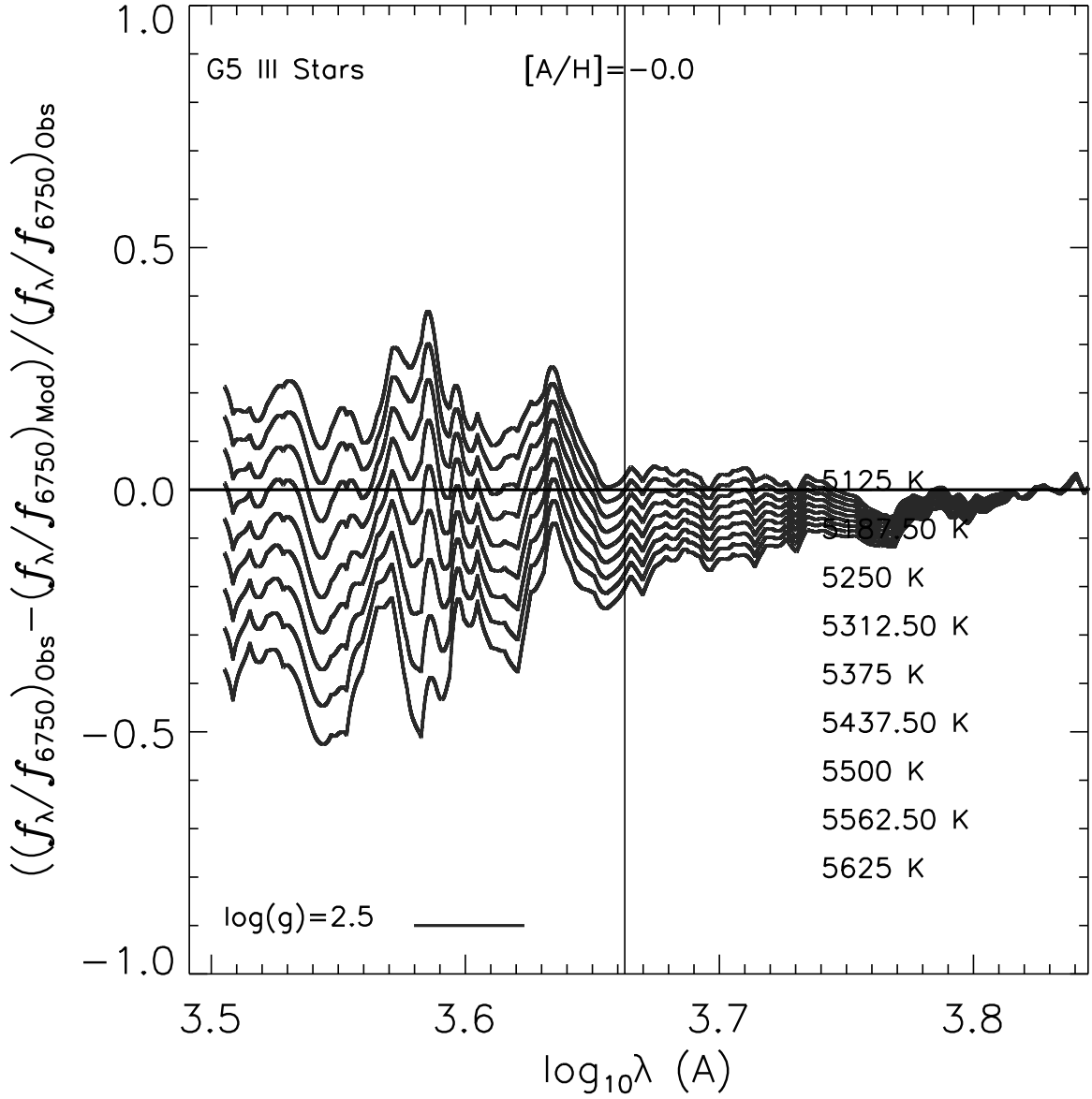}
\caption{G5 III sample - Spectra of the relative difference between the observed sample average 
$f_{\lambda, {\rm 6750}}$ spectrum and synthetic $f_{\lambda, {\rm 6750}}$ spectra of Series 1 models.  The horizontal line
indicates a difference of zero.  The vertical line represent the break-point between the ``blue'' and
``red'' bands. 
  \label{fdiffG5III}}
\end{figure}

\begin{figure}
%\plotone{K3-4IIIStars.diff.eps}
\plotone{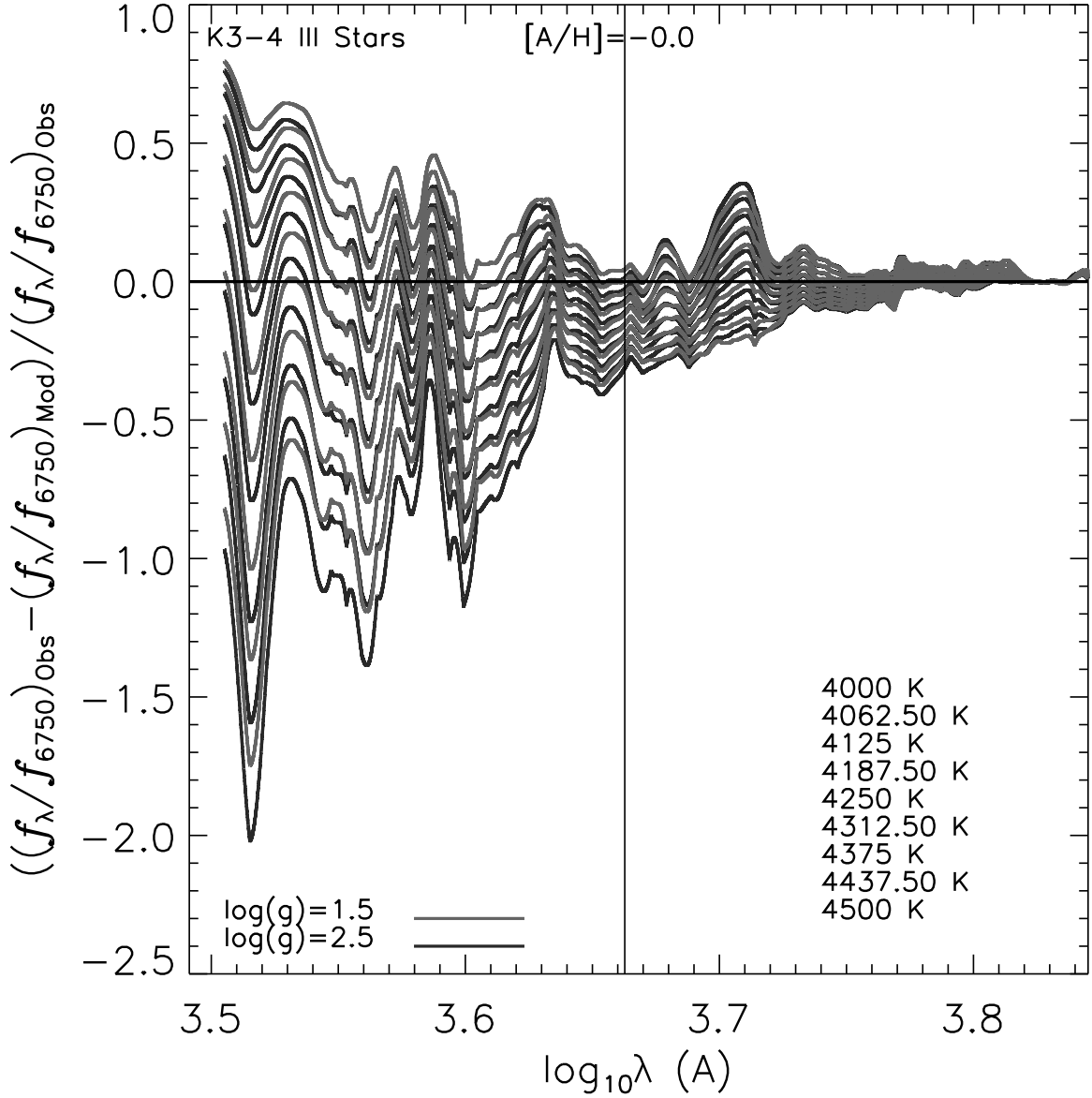}
\caption{Same as Fig. \ref{fdiffG5III}, but for the K3-4 sample.  Black lines: $\log g=2.5$, gray
lines: $\log g=1.5$.  For clarity, the synthetic spectra the $\log g=2.0$ models have been omitted.  
  \label{fdiffK4III}}
\end{figure}

All spectra, observed and synthetic, have been normalized by dividing 
by the average of their flux in the 6600 to 6900 \AA~ region, chosen to be just blue 
of the first significant telluric contamination bands due to O$_{\rm 2}$ and 
H$_{\rm 2}$O, to produce the distribution 
denoted $f_{\lambda, {\rm 6750}} = {f_\lambda\over f_\lambda(\lambda={\rm 6750})}$.
Figs. \ref{fcompG5III} and \ref{fcompK4III} show the comparison of the mean and
$\pm 1 \sigma$ spectra of the observed $f_{\lambda, {\rm 6750}}$ distributions with 
the closest matching and bracketing
 synthetic $f_{\lambda, {\rm 6750}}$ distributions for the hottest and coolest 
giant samples.  
Figs. \ref{fdiffG5III} and \ref{fdiffK4III} show the difference
between the mean of the observed $f_{\lambda, {\rm 6750}}$ distribution and the closest matching 
and bracketing synthetic distributions relative to the observed mean distribution,
$( f_{\lambda, {\rm 6750, Mean Observed}} - f_{\lambda, {\rm 6750, Model}} ) / f_{\lambda, {\rm 6750, Mean Observed}}$ 
for the same two samples.

\section{Goodness of fit statistics \label{fitstats}}

 We interpolate all observed and synthetic spectra onto a uniform $\lambda$ grid of $\Delta\lambda=0.01$ \AA~ that moderately 
over-samples 
all spectra at all $\lambda$ values, and compute for each spectral class sample the root mean square {\it relative} 
deviation, $\sigma$, of the mean observed $f_{\lambda, {\rm 6750}}$ distribution from
the closest matching and bracketing synthetic $f_{\lambda, {\rm 6750}}$ 
distributions in the $\lambda$ 
range from 3200 to 7000 \AA, according to\\

\begin{equation}
\sigma^2 = {1\over N}\sum_i^N ((f_{\lambda, {\rm 6750, Obs}}-f_{\lambda, {\rm 6750, Mod}})/f_{\lambda, {\rm 6750, Obs}})^2
\end{equation}

 where $N$ is the number of $\lambda$ points in the $\lambda$ grid in the 3200 to 7000 \AA~ range.  
 
\paragraph{}

We also compute separate RMS values, $\sigma_{\rm blue}$ and $\sigma_{\rm red}$, for the 
so-called ``blue'' and ``red'' sub-ranges
of 3200 to 4600 \AA~ and 4600 to 7000 \AA, respectively.   
We also compute the mean relative deviations, $\Delta_{\rm blue}$ and $\Delta_{\rm red}$, for the
two sub-ranges, according to. 

\begin{equation}
\Delta = {1\over N}\sum_i^N (f_{\lambda, {\rm 6750, Obs}}-f_{\lambda, {\rm 6750, Mod}})/f_{\lambda, {\rm 6750, Obs}}  
\end{equation}
 
A comparison of
the blue and red $\sigma$ values indicates how well the synthetic
spectra fit in the blue and near UV band given the quality of fit in the red band.
The comparison of $\Delta$ between the bands allows an assessment of systematic
discrepancies throughout the blue compared to the red because, unlike $\sigma$,
$\Delta$ retains its sign.   
A break-point of 4600 \AA~ was chosen on the basis of visual inspection of 
where the deviation of the synthetic from the observed spectrum starts to become rapidly larger as $\lambda$ decreases.
In Tables \ref{tabstats1} and \ref{tabstats2} we present the $\sigma$, $\sigma_{\rm blue}$, 
and $\sigma_{\rm red}$ values for the Series 1 and 2 models, respectively, along with the
best fit value of $T_{\rm eff}$ and $\log g$ for each star.  The value of the model 
$[{{\rm A}\over{\rm H}}]$ is also tabulated, although, its value was specified {\it a priori}
 rather than fitted.  

\subsection{Trend with $T_{\rm eff}$}

\begin{figure}
%\plotone{fit-solgiants.eps}
\plotone{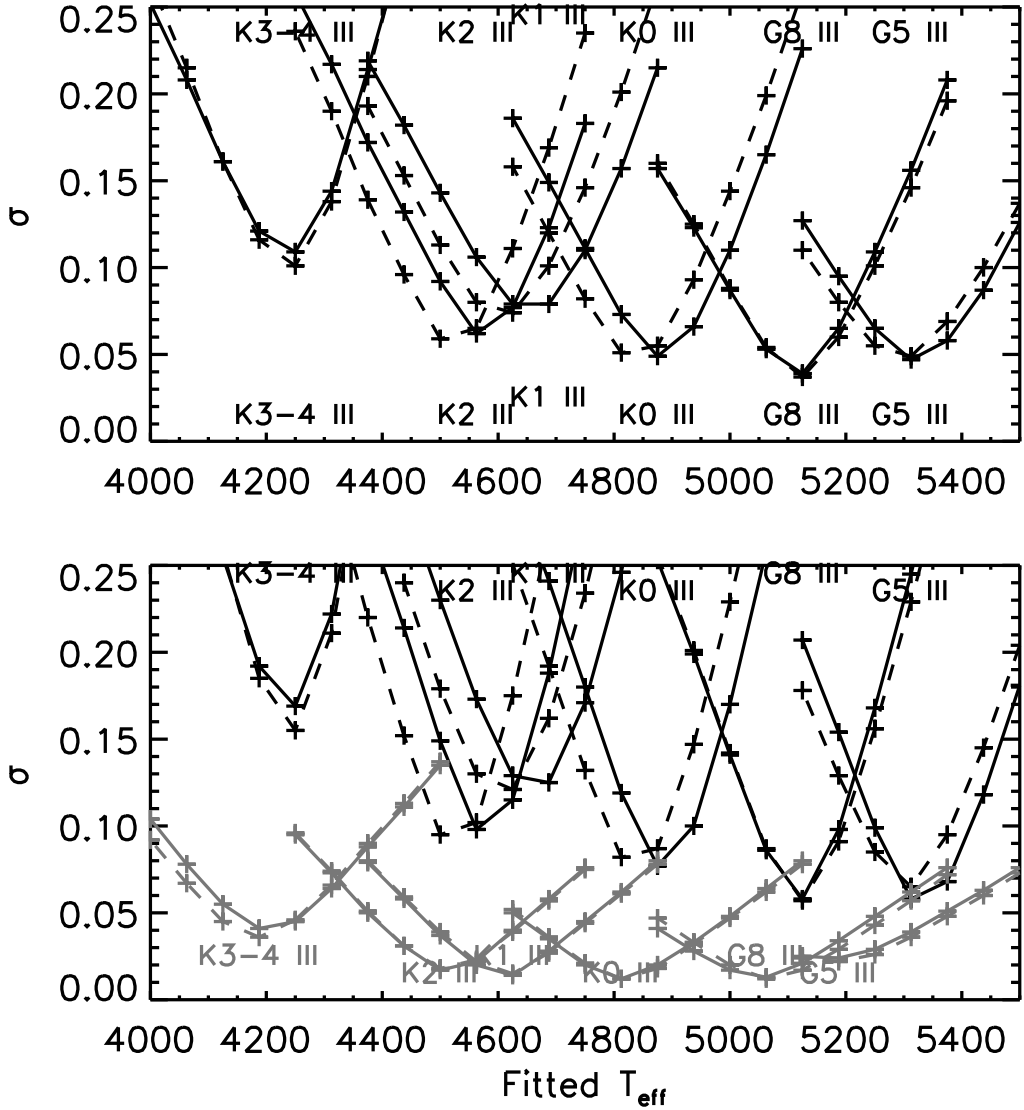}
\caption{Giants of solar metallicity: Variation of $\sigma$ with model $T_{\rm eff}$.   
Solid line: Series 1 models; dashed line: Series 2 models.  Upper panel: Fit to total SED;
 Lower panel: Black lines: fit to blue band; gray lines: fit to red band.  \label{fstatsgnt}}
\end{figure}

Fig. \ref{fstatsgnt} shows the variation of
$\sigma$, $\sigma{\rm blue}$, and $\sigma_{\rm red}$ with model $T_{\rm eff}$
for the giant stars of solar metallicity.  The best fit $\sigma$ value generally increases with 
increasing lateness of the spectral class. 
We note that the density of spectral lines generally increases with increasing lateness.
Therefore, this trend in the discrepancy between synthetic and observed SEDs could be
explained by inadequacies is the input atomic data for bound-bound ($b-b$) transitions,
or by inadequacies in the treatment of spectral line formation.  The best fit $\sigma$ values
for the Series 1 and 2 models differ negligibly for the earlier spectral classes, and
there is marginal evidence that the Series 2 models provide a slightly better fit 
(lower $\sigma$ value) for the latest spectral classes. 

\subsubsection{G5 III sample}

The behavior of the variation of $\sigma_{\rm red}$ with $T_{\rm eff}$ for the G5 III stars
is peculiar and leads to a spurious result for the best fitted value of $T_{\rm eff}$.  From
Fig. \ref{fdiffG5III} it can be seen that this is caused by a broad absorption feature exhibited
by the observed SED with respect to the model SEDs ranging from a $\log\lambda$ value of 3.753 to 3.774
(5660 to 5940 \AA).  As a result, the value of $\sigma_{\rm red}$ is increased significantly,
even for models that provide a good match to the overall spectrum.  Therefore, our
best fit value of $T_{\rm eff}$ for the G5 III models is best determined from 
the blue band alone.  This deficit of absorption in the synthetic SEDs with respect to the
observed ones is consistently present in the individual observed spectra for
the G5 III stars, spans 12 data points in the raw observed spectrum, and varies smoothly with
wavelength over a range of ~280 \AA.  Therefore, it is likely caused by a cluster of spectral
lines that are either missing, or are too weak, in the model spectra.  
  We note that this
discrepancy is either absent, or much less pronounced, in both the G4-5 V and G8 III stars, so
appears to be localized in both $T_{\rm eff}$ and $\log g$.  

\paragraph{}

We have examined histograms of the average numbers of spectral lines per \AA~ 
by atomic chemical species in a synthetic spectrum of a model
of $T_{\rm eff}$/$\log g$/[${{\rm A}\over{\rm H}}$]=5250/2.5/0.0, representative of the models
that fit the observed G5 III SED.  We compared three 
$\lambda$ ranges: the problematic 5660 to 5940 \AA~ range, and the bracketing ranges of
5000 to 5660 and 5940 to 7000 \AA.  All three ranges show approximately the 
same pattern and same absolute average numbers of lines per unit wavelength for
all atomic species that contribute a significant number of lines.  We are
unable to identify any suspect chemical species for which there is an excess
or dearth of lines in the 5660 to 5940 region compared to neighboring regions 
that might provide a clue to the cause of the discrepancy.    

\paragraph{}

\begin{figure}
%\plotone{jacobicomp.eps}
\plotone{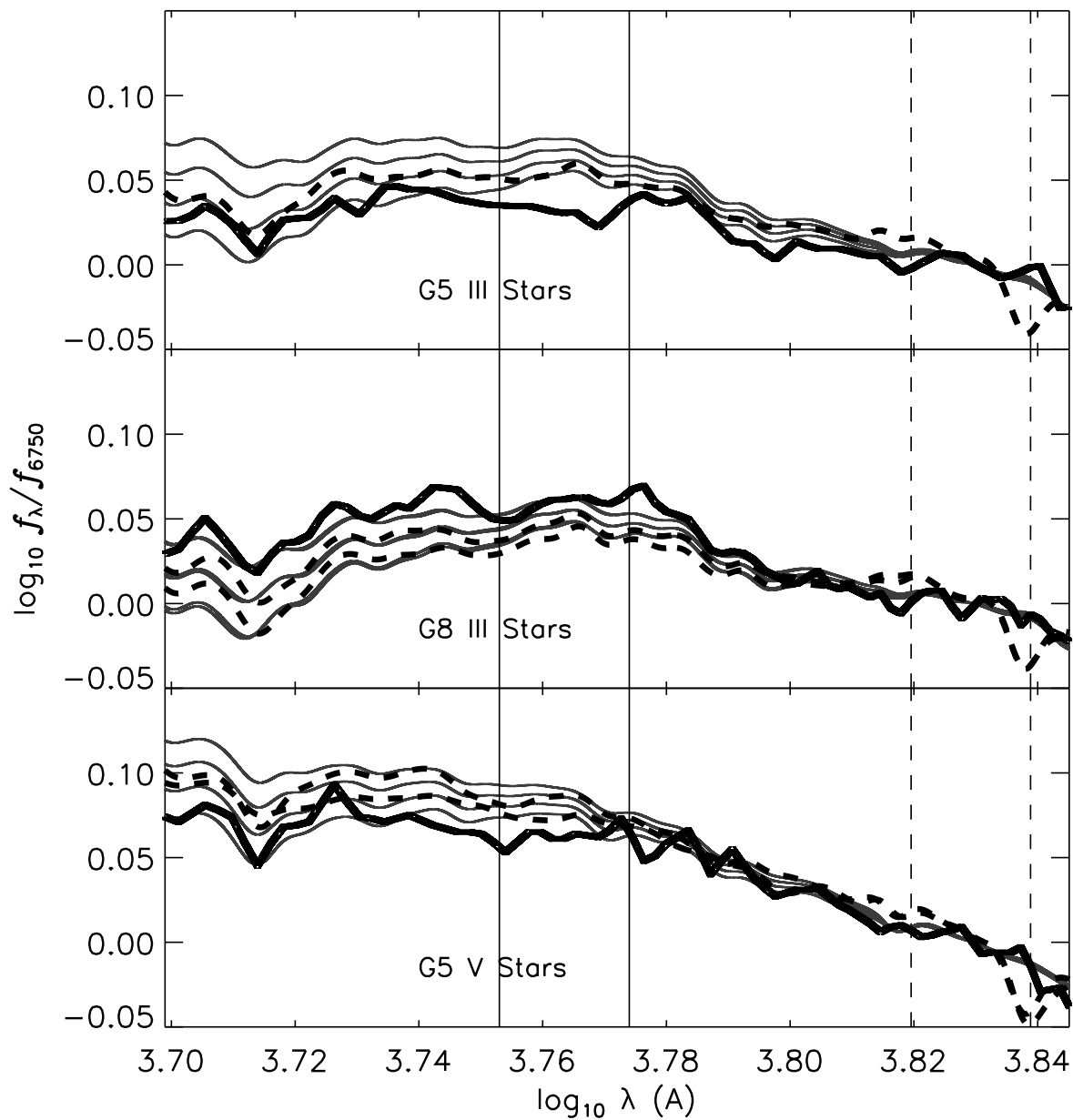}
\caption{Comparison of B85 and \citet{jacobyhc84} observed spectra.  Solid black line: sample mean from 
the B85 catalog as in Fig. \ref{fvaryG5III}. solid gray lines: synthetic spectra as in Figs. 
\ref{fcompG5III} and \ref{fcompK4III}, dashed black lines: spectra from library of \citet{jacobyhc84} for
a G5 III star (BD+281885, upper panel), a G7 (HD 249240) and a G8 (HD 245389) III star (middle panel),
and a G4 (TR A 14) and a G6 (HD 22193) V star (lower panel).  Vertical lines: solid: region of discrepancy
between the B85 and PHOENIX spectra for G5 III stars; dashed: normalization region. \label{jacoby} }       
\end{figure}

We have compared our mean observed spectra based on the B85 catalog for the G5 III and V, 
and G8 III types to representative spectra from the stellar spectrophotometric library of \citet{jacobyhc84}.
In Fig. \ref{jacoby} we show the comparison in the $\lambda$ range around the region of the
G5 III discrepancy.  The
\citet{jacobyhc84} library does not have a spectrum for type G5 V, so we have compared our 
spectrum of that class to spectra for classes G4 and 6 V.  The G4 V star is designated
'TR A 14' and \citet{jacobyhc84} flag it as a star that could not be identified.  However,
for lack of an alternative, we use it as a comparison.  There are small systematic
differences between our mean B85 spectra and those of \citet{jacobyhc84} over a broad range of 
$\lambda$ for all three spectral types.  However, for the G5 III sample, the \citet{jacobyhc84}
spectrum is distinctly brighter than our mean B85 spectrum in the $\log\lambda$ 3.753 to 3.774 range,
as compared the the bracketing $\log\lambda$ ranges, and is in greater agreement with our
synthetic SED.  We conclude that there may have been a data acquisition or calibration problem
with the spectra in the B85 catalog that is very particular to the G5 III class.  From Fig.
\ref{jacoby} it can be seen that the B85 and \citet{jacobyhc84} spectra are much more consistent 
with each other throughout this region for the G5 V and G8 III classes.  We note again that
our B85 G5 III sample suffers from small-number statistics in that it consists of three spectra of two 
stars.  In what we follows we only draw conclusions for the blue spectral band of the G5 III sample. 
(A full systematic comparison of our B85 mean spectra with those \citet{jacobyhc84} throughout our
$\lambda$ range is beyond the scope of this investigation, but, the comparison in Fig. \ref{jacoby}
is generally encouraging.)

\subsubsection{Red {\it vs} blue band}   

For all samples, the best fit to the red band has a $\sigma$ value that is lower than that
of the fit to the blue band by 0.05 to 0.1.  This may partly reflect that all spectra 
were normalized to a common relative flux value in the red band (6750 \AA).  However, 
from Figs. \ref{fdiffG5III} and \ref{fdiffK4III} it can be seen that the difference
spectra show increasing variability around the zero line in addition to any systematic
trend away from the zero line.  This indicates that the quality of the fit to the mean
 observed spectrum worsens with decreasing $\lambda$ regardless of how the observed and
synthetic spectra were normalized. 
For all solar metallicity giants, the Series 1 models fitted to the blue band consistently
give best fit $T_{\rm eff}$ values that are one $\Delta T_{\rm eff}$ resolution element, 63K,
higher than those fitted to the red.  For the Series 2 models, both bands yield the same
best fit $T_{\rm eff}$ value for the K0 to K2 III stars, whereas for the two ends of the
$T_{\rm eff}$ range where both bands can be used, G8 and K3-4, the Series 2 models lead
to the same pattern as the Series 1 modes, with the blue band fit yielding a higher $T_{\rm eff}$ 
value.  This may provide marginal evidence that the Series 2 models lead to a greater
consistency of fit throughout the SED. 

\subsubsection{Series 1 {\it vs} Series 2}

For the samples at the two ends of our $T_{\rm eff}$ range, G5 to G8 and K3-4, fits with 
the Series 1 and 2 models both lead to the same value of inferred $T_{\rm eff}$, to within the $\Delta T_{\rm eff}$
precision of the grid.  However, for the intermediate samples, K0 to K2, the Series 2 models
fitted to the blue and to the overall SED consistently lead to fitted $T_{\rm eff}$ values
that are one $\Delta T_{\rm eff}$ resolution element lower than that of the Series 1
models.  Because the Series 2 models have less line blanketing, they predict greater 
flux in the blue with respect to the red than do the Series 1 models.  Therefore, we expect
them to yield lower fitted $T_{\rm eff}$ values to a given observed SED. 

\subsection{Trend with $\log g$}

\begin{figure}
%\plotone{fit-dwarfs.eps}
\plotone{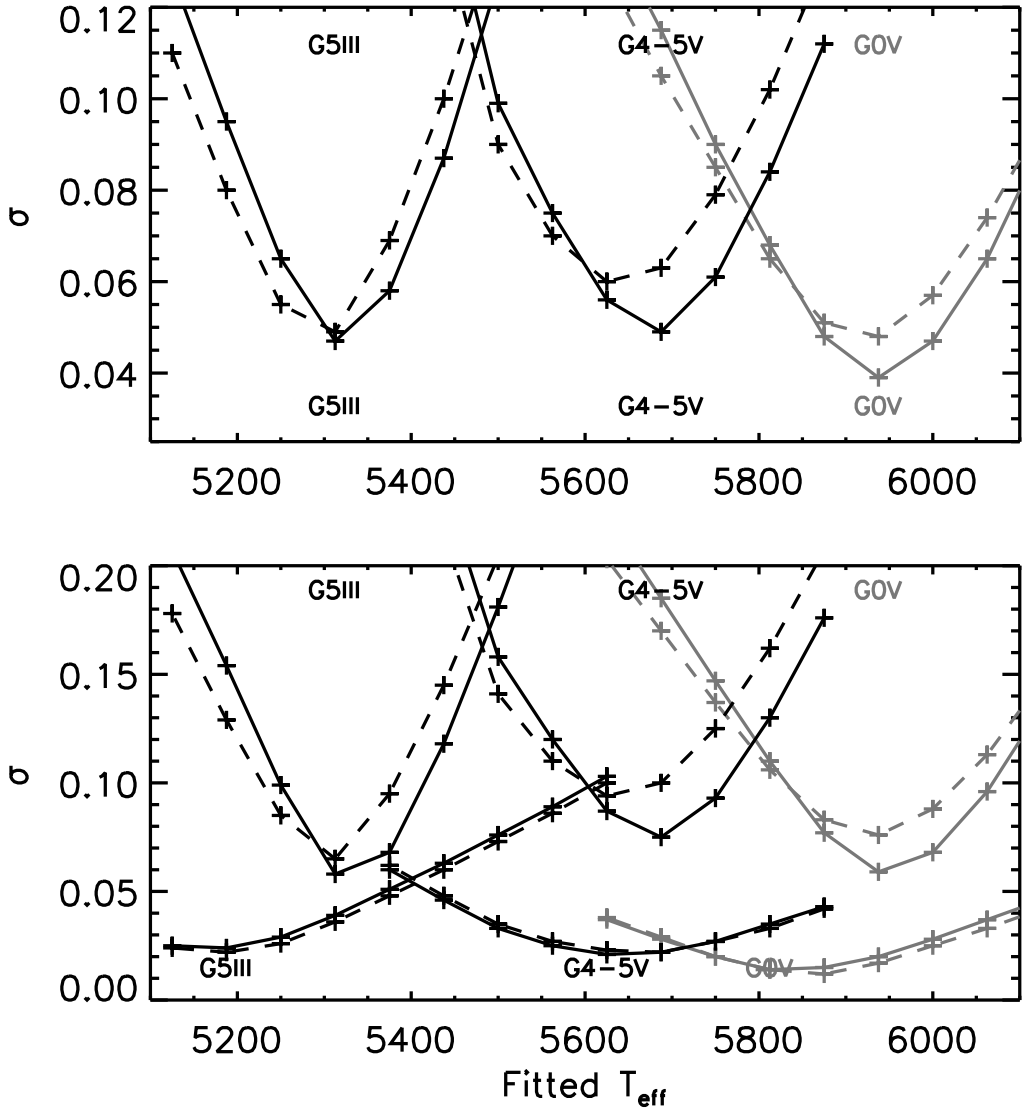}
\caption{Same as Fig. \ref{fstatsgnt}, but for the dwarf stars.
  \label{fstatsdwf}}
\end{figure}

Fig. \ref{fstatsdwf} shows the fitting quantities for the 
G5 III, G4-5 V, and G0 V stars of solar metallicity.  Examination of these figures allows a 
limited assessment of how the
quality of fit varies in the $\log g$ dimension at spectral class G5.
The Series 1 models provide fits of similar quality to the total SED of G5 stars of both 
luminosity
classes, III and V.  By contrast, the Series 2 models provide a significantly worse fit 
to the class V than to the class III stars, with the quality of fit to class III differing 
negligibly from that provided by the Series 1 models.  Furthermore, we note that 
this same discrepancy in the quality of fit provided by Series 1 and 2 is also seen
for the G0 V stars.  Fig. \ref{fstatsdwf} shows that these trends in the quality
of the fit to the total band is driven mainly by the quality of fit to the blue band.
 The suggestion is that the models computed with the ``big'' input line 
list provide a better fit to the blue band spectra of dwarfs than, and as good a fit to the giants as, 
the models computed with the ``small'' line list. 

\paragraph{}

For the G4-5 V stars, the $T_{\rm eff}$ value fitted to the blue band with Series 1 models 
is 63 K higher than that fitted with Series 2 models, whereas, for the G 5 III stars, 
both Series yield the same blue band $T_{\rm eff}$ value.  The G4-5 V stars are anomalous 
among our solar metallicity stars in that the Series 2 models give a {\it higher} fitted 
$T_{\rm eff}$ value to the red band than to the blue band. 
Unfortunately, our ability to compare the fit to the red band as a function of
$\log g$ is undermined by the peculiarities with modeling the red band of the G5 III sample
discussed earlier in this section.  
%said above? For the blue band and over all SED, the Series 1 and 2 models 
%said above? lead to the same best fit 
%said above? $T_{\rm eff}$ for the G0 V stars, whereas for G5 V, the Series 2 models yield a $T_{\rm eff}$
%said above? value that is 63 K lower than that of Series 1.

\subsection{Trend with [${{\rm A}\over{\rm H}}$]}

\begin{figure}
%\plotone{fit-metalpoor.eps}
\plotone{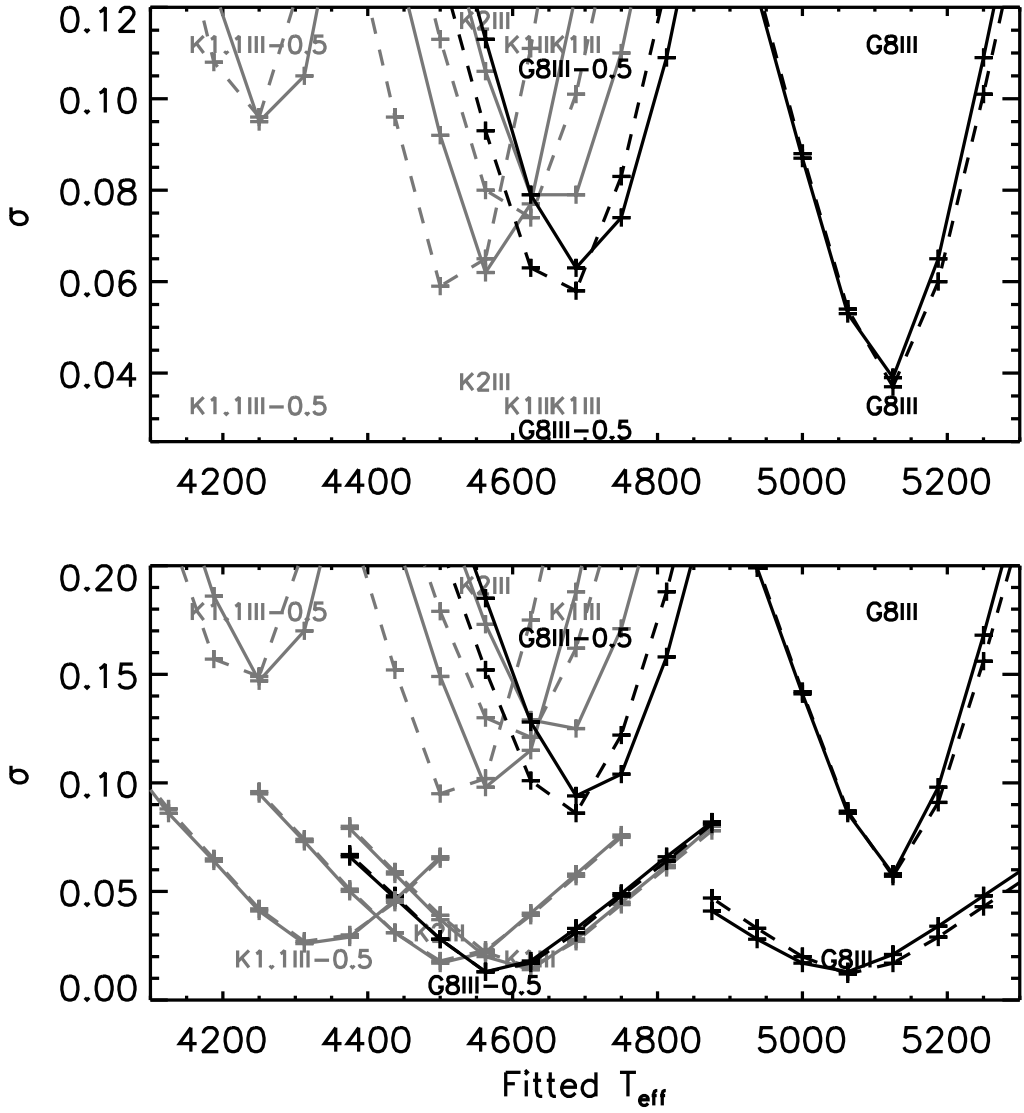}
\caption{Same as Fig. \ref{fstatsgnt}, but for the metal poor giants.
  \label{fstatsmtl}}
\end{figure}

Fig. \ref{fstatsmtl} shows the same quantities for the
G8 III and K1 to K2 III  stars of [${{\rm A}\over{\rm H}}$] equal to 0.0 and -0.5, and
allows a limited assessment of how the quality of fit varies in the metallicity dimension. 
For a given {\it spectral class}, both the Series 1 and 2 models provide a 
significantly better quality of fit to the stars of solar metallicity than to
those that are metal poor.  This may in part reflect the inappropriateness of 
scaled solar abundance models for fitting the population of stars of 
[${{\rm A}\over{\rm H}}$]$=-0.5$.  By contrast, for a given $T_{\rm eff}$ value, 
there is some evidence that all models provide a similar quality of fit, independent of 
[${{\rm A}\over{\rm H}}$].
We note that the metal-poor G8 III star has a 
best fit $T_{\rm eff}$ value closer to that of the solar metallicity K1 to K2 stars than
that of the solar metallicity G8 III star, and the
value of $\sigma$ for the red and blue bands, and for the overall SED, is similar for
the G8 III/[${{\rm A}\over{\rm H}}$]$=-0.5$ and K2 III/[${{\rm A}\over{\rm H}}$]$=0.0$ stars.  
This suggests that the strongest relation is the anti-correlation between
$T_{\rm eff}$ and $\sigma$ rather than that between [${{\rm A}\over{\rm H}}$] and $\sigma$.    
%True? The trend with [${{\rm A}\over{\rm H}}$]$=-0.5$ is echoed by 
%True? the quality of fit to the blue spectrum much more than the red, and we conclude that
%True? the quality of fit trend to the overall SED is being driven by the blue band,  This 
%True? is to be expected because the treatment of line extinction is more critical
%True? to fitting the more heavily line blanketed blue band.  

\subsection{Arcturus}

  For the Arcturus sample (K1.5 III, [${{\rm A}\over{\rm H}}$]$=-0.5$) the $T_{\rm eff}$ value
fitted to the red band is higher than that fitted to the blue by 63 K.  Therefore,
models fitted to the red band would predict too much flux in the blue and near UV bands.  This is
consistent with the results of \citet{shorth03} and \citet{shorth09}, who also compared
models to the observed $f_\lambda$ distribution of B85.  We note that 
62.5 K is the numerical precision of
our $T_{\rm eff}$ grid.  Therefore, the most we can conclude is that the discrepancy
in the best fit $T_{\rm eff}$ value between the red and blue bands is in the range of 
31 to 125 K.  By contrast, the solar
metallicity giants of spectral type G8 to K4 do not show this trend, having best fit
$T_{\rm eff}$ values that are {\it lower} by 63 K in the red band (to within the precision of 
our model grid).  Moreover, our only other 
metal poor sample, G8 III, also shows the opposite trend as Arcturus, with a red band
$T_{\rm eff}$ value that is 125 K (two $\Delta T_{\rm eff}$ increments) lower than that of 
the blue band.  The suggestion is that Arcturus may be peculiar among giants generally in 
that models fitted to the red over-predict the blue band flux.

\section{Comparison to other $T_{\rm eff}$ calibrations}

\subsection{Empirical calibration}

\begin{figure}
%\plotone{calibrate-solgiants.eps}
\plotone{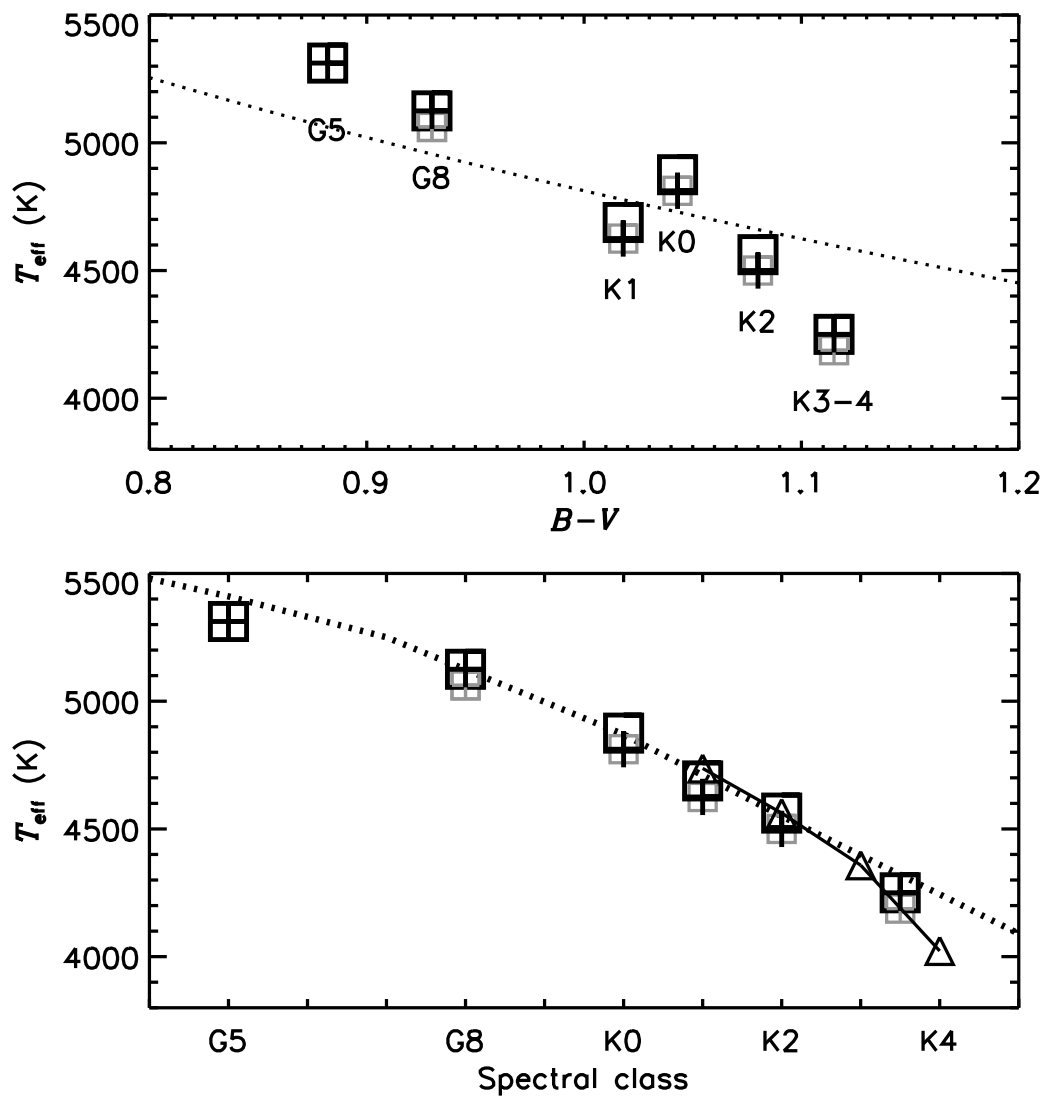}
\caption{Comparison of our best fit $T_{\rm eff}$ values with other $T_{\rm eff}$ calibrations.
Squares: Series 1 models; Crosses: Series 2 models.  Black symbols: Fit to the blue band; Gray symbols: 
fit to the red band.  Dotted lines: Upper panel: Empirical calibration of RM05; Lower panel: 
PHOENIX NextGen models fitted to stellar spectral libraries (BBCR04).  Triangles (lower panel):
$T_{\rm eff}$ values of B10. 
  \label{fcalibgnt}}
\end{figure}

\begin{figure}
%\plotone{calibrate-dwarfs.eps}
\plotone{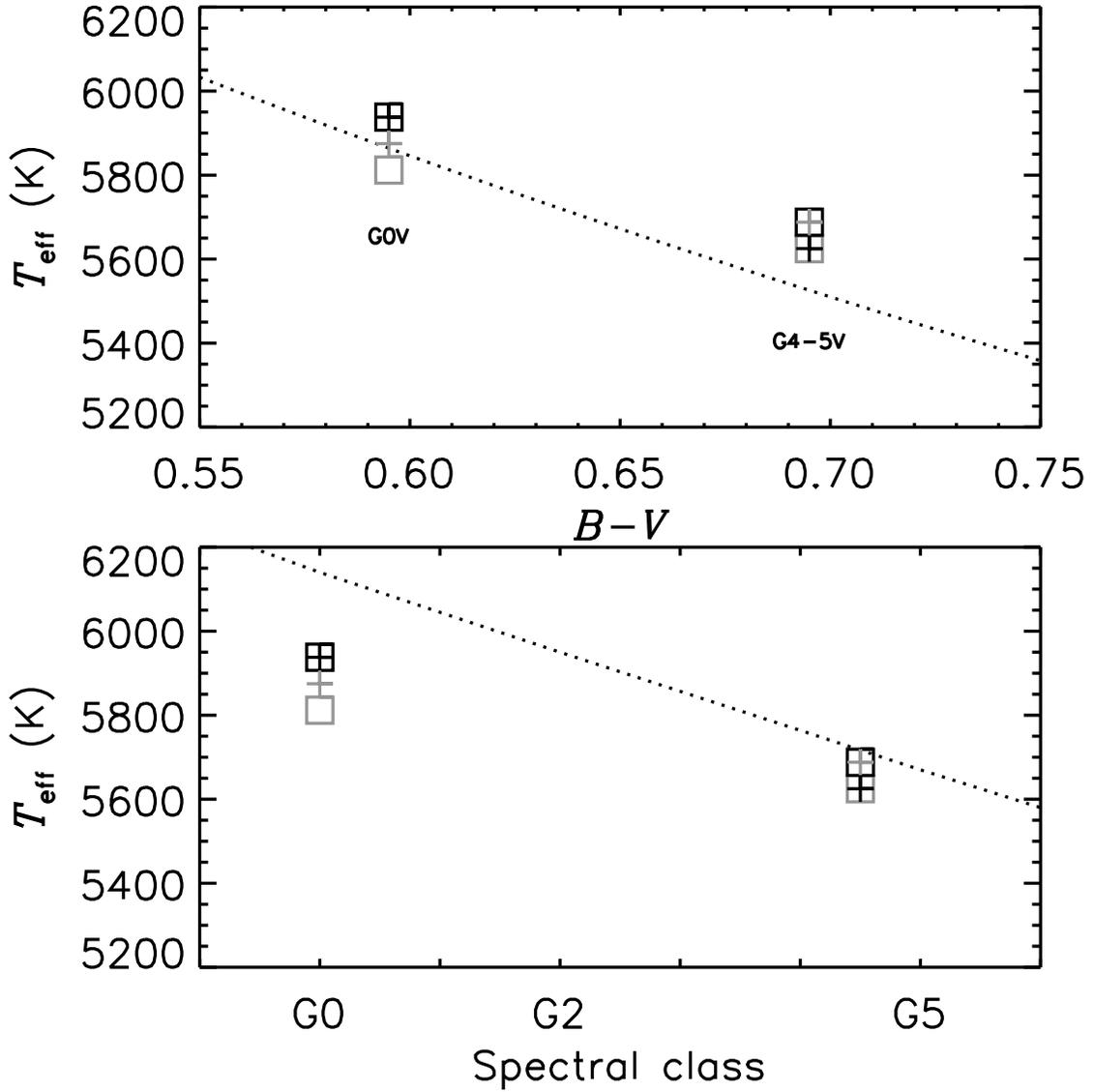}
\caption{Same as Fig. \ref{fcalibgnt}, but for the dwarfs. \label{fcalibdwf}}
\end{figure}

\begin{figure}
%\plotone{calibrate-metalpoor.eps}
\plotone{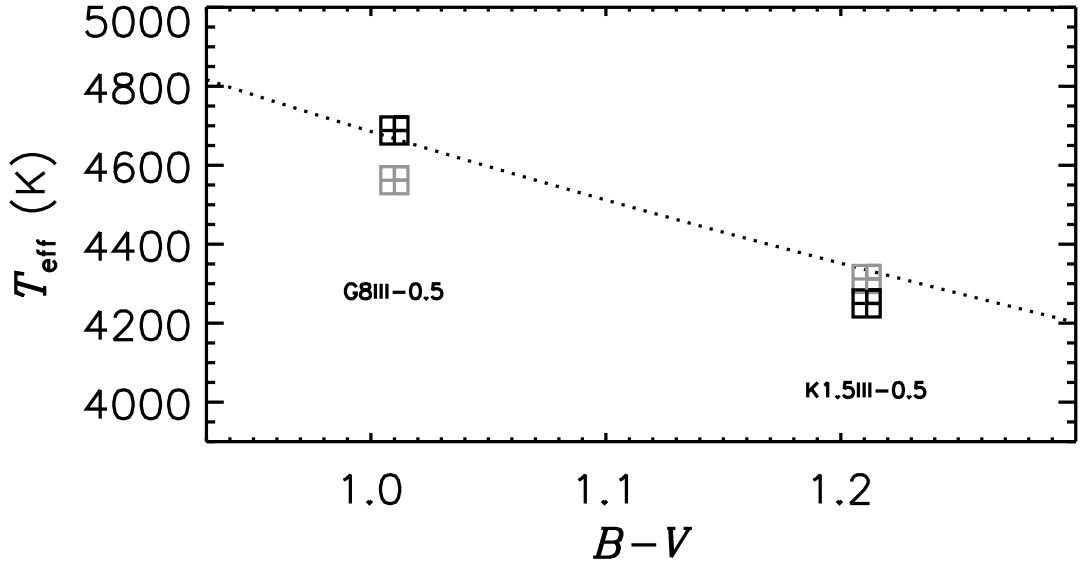}
\caption{Same as Fig. \ref{fcalibmtl}, but for the metal poor stars.  Note that
BBCR04 only analyzed solar metallicity stars, so we only show a comparison to
the $T_{\rm eff}$ calibration of RM05.  \label{fcalibmtl}}
\end{figure}

\citet{ramirezm05} (RM05) present an empirical determination of $T_{\rm eff}$ based
on applying the Infrared Flux Method (IRFM) to 100 
dwarfs and giants of spectral class from F0 to K5 and $[{{\rm A}\over{\rm H}}]$
from -4 to +0.4 as a function of many observables, including Johnson $B-V$
color.  This work is an extension of the definitive $T_{\rm eff}$ calibration
work of \citet{alonsoam99} and other papers in that series.  They quote standard 
deviations of 30 to 120 K, and find that their IRFM $T_{\rm eff}$ scale 
agrees to within 10 K with directly determined values for stars with
measured angular diameters.  RM05 do not present $T_{\rm eff}$
as a function of spectral class.  However, we have computed mean and RMS ($\sigma$)
$B-V$ values for each of our spectral class samples using colors for individual objects 
from the Catalogue of Homogeneous Means in the UBV System \citep{mermilliod91}.  
We then used the fitted Eqs., 1 and 2, 
and the fitting co-efficients of Tables 1 and 2, 
of RM05 to interpolate in $B-V$ and produce empirical 
$T_{\rm eff}$ values for each of our spectral class samples and metallicities. 
We checked our interpolated $T_{\rm eff}$ values against, presumably, less accurate values found 
from linear interpolation in Tables 4 and 5 of RM05 and found them to
be consistent.  In Table \ref{tabcomp1} and Figs. \ref{fcalibgnt} through \ref{fcalibmtl} we 
present a comparison of our 
$T_{\rm eff}$ values fitted to our blue and red spectral ranges, and those of the
RM05 calibration.  Because of the peculiarities of fitting the red band of the
G5 III stars described in section \ref{fitstats}, the red band $T_{\rm eff}$ value has
been suppressed in Figs. \ref{fcalibgnt} through \ref{fcalibmtl}. 

\paragraph{}

\citet{baines10} (B10) used the CHARA array to interferometrically measure the angular diameters of 
25 K giants in the $K$ band, including eight
in the spectral class range K0 to K4 III with $[{{\rm A}\over{\rm H}}]$ values within $\pm 0.1$ of
0.0, and two in the K1 to 2 III range with $[{{\rm A}\over{\rm H}}]$ values within $\pm 0.1$ of
-0.5.  They combined their angular diameter measurements with distances from Hipparcos and
 photometrically inferred bolometeric flux values to derive $T_{\rm eff}$ values.  We have 
calculated averages for their $T_{\rm eff}$ values for solar metallicity giants of spectral
class K1 (three stars), K2 (two stars), K3 (1 star) and K4 (2 stars), and include these values
in Table \ref{tabcomp1} and Fig. \ref{fcalibgnt}.  They also present a 
measured $T_{\rm eff}$ value for an Arcturus analog (HD 170693, K1.5III-0.5).  These results were
published just as we were completing our investigation, so we included them for comparison.  
Unfortunately, none of the B10 stars are in the samples we selected from the B85 catalog.

\subsubsection{Solar metallicity giants}

For the G8 to K1 stars, the models of both series fitted
to the red band give $T_{\rm eff}$ values closest to the RM05 values, and are 
consistently higher than RM05
by less than 100 K.  For the K2 sample, our red
band fit matches the RM05 value to within the $\pm 62.5$ K precision of our grid.  
The Series 1 models fitted to the blue band consistently yield $T_{\rm eff}$ values
that are higher than the RM05 $T_{\rm eff}$ calibration.
An equivalent alternate interpretation is that the Series 1 models
with the RM05 $T_{\rm eff}$ value for their observed $B-V$ color would predict
too {\it little} blue band flux compared to the observed SED, while providing a
closer match to the red band.  

\paragraph{}

The B10 $T_{\rm eff}$ values for classes K1 and 2 III are is closest agreement with our Series 1
values derived from the blue band, being an exact match at K2 III to within the
precision of the respective values.  Their value for the K3-4 stars is lower than our lowest
value by $\approx 50$ K.

%Huh??
%\paragraph{}
%
%There is an anomaly at spectral class K0 III where models of both Series fitted to the red
%band yield lower $T_{\rm eff}$ values than those fitted to the blue band, by contrast with the
%behavior at the adjacent spectral classes.  However, the models fitted to the K0 III sub-sample
%with ``low'' blue SEDs have derived $T_{\rm eff}$ values that are 63 K
%lower than the blue band $T_{\rm eff}$ value of the whole sample, and would be
%in closer agreement to the red band $T_{\rm eff}$ value.  Therefore, the K0
%anomaly may be related to the bifurcation in the distribution of blue band SEDs
%for the K0 III sample.  

\subsubsection{Solar metallicity dwarfs}

 For the G0 V stars, the Series 2 models fitted to the red band give a $T_{\rm eff}$ value very close
to the RM05 value.  All models fitted to the blue band give $T_{\rm eff}$ values that are too high by
$\approx 100$ K.
For the G4-5 stars all models give $T_{\rm eff}$ values that are consistent with each other, but 100 to 
150 K higher than that of RM05.  These results are similar to those found for G5 III stars in the
blue band, so
the discrepancy between our blue band $T_{\rm eff}$ scale and that of RM05 at spectral class G5  seems 
to be independent of $\log g$ in the 2.5 to 4.5 range,  

\subsubsection{Metal poor giants}

For the G8 III stars of $[{{\rm A}\over{\rm H}}]=-0.5$, the Series 1 and 2 models
fitted to the blue band provide a very close match to the $T_{\rm eff}$ value of RM05, better 
than the fit to the solar metallicity G8 giants.  All models fitted to the red band provide $T_{\rm eff}$ 
values that are 100 to 150 K lower than that of RM05.  For Arcturus (K1.5III-0.5), all models fitted
to the red band give $T_{\rm eff}$ values in close agreement with RM05, being lower by 20 K.  This
result is consistent with our result for solar metallicity K1 and 2 stars, and suggests that the 
quality of our fit to the red band for early K giants is independent of metallicity in this
[${{\rm A}\over{\rm H}}$] range.  The B10 value for their one K1.5 III star of $[{{\rm A}\over{\rm H}}]=-0.5$
(HD 170693) is 75 K higher than our highest value.  
%Old?? For the blue band, all models give $T_{\rm eff}$ values that are lower than RM05 by $\approx 100$ K.

\subsection{Previous PHOENIX calibration}

\citet{bertonebcr04} (BBCR04) derived $T_{\rm eff}$ values from fitting theoretical
SEDs for models of $[{{\rm A}\over{\rm H}}]=0.0$ taken from the NextGen grid
computed with an earlier version of PHOENIX (V10) \citep{hauschildtafba99}, 
The most important differences between V10 and V15 that are relevant to these
models are 1) Improvements in the treatment of the line profiles so that many weak
to moderate strength lines that were treated as Gaussian are now treated with
Voigt profiles, and 2) Improvement in the quality of a wide variety of EOS input data. 
\citep{hauschildtab99} to SEDs of solar metallicity 
dwarfs and giants of spectral class from early A to mid-M taken from the  
observational spectral libraries of \citet{gunns83} and \citet{jacobyhc84}. 
We note that the NextGen models were computed using the ``big'' atomic line list of our
Series 1 models.  Among 
the main differences between the NextGen grid and the one presented here are: 1) the 
NextGen
models of $\log g>3.5$ are computed with plane-parallel geometry, however,
the effects of sphericity are expected to be negligible for that surface gravity
range. 2) the NextGen models
have the value of $\xi_{\rm T}$ set to 2 km s$^{-1}$ throughout the grid, independent
of $\log g$.  As discussed above, our value of $\xi_{\rm T}$ decreases to 
1 km s$^{-1}$ for the dwarfs on the basis of our experience modeling the disk 
integrated flux spectrum of the Sun \citep{shorth05}.  BBCR04 do not provide an interpolation formula for
$T_{\rm eff}$, so we have interpolated linearly in their Table 1 to find $T_{\rm eff}$
values for our spectral classes that are missing from their table.
Table \ref{tabcomp1} and Figs. \ref{fcalibgnt} through \ref{fcalibmtl} include a comparison the 
$T_{\rm eff}$
values of BBCR04.

\subsubsection{Solar metallicity giants}

Generally, the BBCR04 $T_{\rm eff}$ scale is 50 to 150 K hotter than the empirical RM05 
scale in our spectral class range,  and our models fitted to the red band consistently provide $T_{\rm eff}$ values 
that are lower than those of BBCR04 by $\approx 50$ K.
We note that our Series 1 models fitted to the blue band provide consistently very good 
agreement to the NextGen $T_{\rm eff}$ scale.  That the Series 1 models would provide better
agreement is expected since they are based on the same atomic line list as the NextGen grid, and
the blue band is expected to show greater sensitivity to the treatment of line opacity
than the red band.
All of our fitted $T_{\rm eff}$ values are lower than the BBCR04 value for the earliest
spectral class, G5 III, which implies that the BBCR04 $T_{\rm eff}$ scale would be even more
discrepant with the empirical RM05 scale than ours is at the hot end of the sequence.  

\subsubsection{Dwarfs and metal poor stars}

Again, for the G4-5 V stars, our Series 1 models fitted to the line-sensitive blue bands provide
the closest match to the NextGen $T_{\rm eff}$ value, which is what we expect.  Somewhat
surprisingly, for the G0V star all our models yield $T_{\rm eff}$ values that are 200 to 250 K
lower than of the NextGen models.  This result
is surprising in that we expect discrepancies due to inconsistencies and inaccuracies in the treatment 
of line blanketing, among other modeling aspects, to be reduced with increasing $T_{\rm eff}$. 
We do not find such a gross discrepancy at G0 V with the empirical RM05 $T_{\rm eff}$ scale.

\section{Conclusions}

 Our giant sequence samples $T_{\rm eff}$ well from spectral class 
G5 to K3-4.  The $\sigma$ value of the best fit to the giant star spectra
 for all spectral 
bands increases with later spectral class, with the value for the 
overall SED increasing from $\approx 0.05$ at spectral class G8 III to 
$\approx 0.10$ at spectral class K3-4 III.  The quality of fit to the
red band (4600 to 7000 \AA) is significantly better than that to the
blue (3200 to 4600 \AA), with the best fit $\sigma$ value being 
$\approx$ 0.02 for classes G8 to K2 III.  (Unfortunately, a peculiar 
discrepancy between the observed and synthetic spectra unique to the
G5 III sample prevents us from meaningfully including that sample in 
conclusions drawn for the red band.)   Higher best fit $\sigma$ values
are driven by an increasing variation around zero in spectra of the difference
between observed and synthetic spectra as $\lambda$ decreases, and 
correlates with the increasing density of spectral lines in $f_\lambda$ 
spectra with decreasing $\lambda$.  Therefore, the contrast between the
best fit $\sigma_{\rm blue}$ and $\sigma_{\rm red}$ values probably
reflects inadequacies in the input atomic line list and/or the treatment of 
line formation.  

\paragraph{}

For the Series 1 models, there is also a {\it systematic}
effect with wavelength in that best fit models to the blue band yield a 
fitted $T_{\rm eff}$ value that is one $\Delta T_{\rm eff}$ grid resolution 
element (63 K) higher than those fit to the red band.  Equivalently, models
fitted to the red band will predict too little flux in the blue band.  This
result is in stark contrast to the results of \citet{shorth03}
and \citet{shorth09}, who found that models fitted to the red band of 
Arcturus (K1.5 III) predicted {\it too much} blue band flux.  We note that 
our Series 1 models fitted to the blue band provide $T_{\rm eff}$ values
that are within 50 K of the values derived with the PHOENIX NextGen grid,
which also used the ``big'' atomic line list.       

\paragraph{}

 The Series 2 models (``small'' line list) fitted to the blue band lead to best 
fit $T_{\rm eff}$ values for the early K giants that are 63 K
lower than do the Series 1 models (``large'' line list), thus providing greater consistency of
best fit $T_{\rm eff}$ value among the spectral bands than do the Series 1 
models.  By contrast, for the two spectral classes for which we have
reliable dwarf spectra, G0 and G4-5, the Series 1 models provide a better fit to 
the blue
band of the dwarfs than those of Series 2, while the two Series provide 
an equally good fit to the red band of dwarfs.  We conclude that there is
marginal evidence that the Series 2 models provide a better fit to solar 
metallicity giants while Series 1 models provide a better fit to solar
metallicity dwarfs.

\paragraph{}

  When comparing giants of the same {\it spectral class}, the models
provide a better fit at solar metallicity than at 
[${{\rm A}\over{\rm H}}$]$=-0.5$,  However, when comparing giants of the same
$T_{\rm eff}$ value, the models give a similar quality of fit 
at both metallicities.  Taking into account the metallicity dependence
in the $T_{\rm eff}$ calibration of the spectral classes, we conclude that
the quality of fit provided by our models is not dependent, or only
weakly dependent, on [${{\rm A}\over{\rm H}}$] in the range -0.5 to 0.0.

\paragraph{}

  We find that Arcturus is peculiar among our stars in that it is the {\it only}
object for which the best fit $T_{\rm eff}$ value to the blue band is {\it lower}
than that for the red band, by 63 K.  This result is qualitatively consistent
with that of \citet{shorth03} and \citet{shorth09}, who found that models
fit to the yellow and red bands of Arcturus yielded significantly more blue 
and near UV band flux than observed.  Those authors concluded that there 
is an important source of near UV band continuous opacity missing from cool
star atmospheric models.  One goal of this work was to determine the extent to 
which this discrepancy was pervasive among late type stars.  We conclude that it
is not, and that Arcturus may be peculiar in this regard.  If anything, there is a 
tendency for models fit to spectra
 of all other spectral classes to {\it under}-predict the blue flux with 
respect to the red. 

\paragraph{}

 For the solar metallicity giants, we find that the Series 2 models, which
yield consistent $T_{\rm eff}$ values between the red and blue bands,
agree to within 100 K with the RM05 empirical $T_{\rm eff}$ values for 
our full range of spectral classes.  For the K2 to K4 III stars, we 
recover the RM05 $T_{\rm eff}$ values to within the $\Delta T_{\rm eff}$ 
resolution of our model grid.  The agreement of the Series 2 $T_{\rm eff}$ 
values with the RM05 values for the G0 and G4-5 V stars is also very good.
For the metal poor giants, the relative success of Series 1 and 2 models is  
mixed with G8 and K1.5 III stars having contrasting results.  
By contrast, the very recently published interferometric $T_{\rm eff}$
values of B10 are in very close agreement to the Series 1 models fit to 
the blue band for early K giants.
We conclude on
balance that the Series 2 models (``small'' atomic line list) provide
greater internal self-consistency and agreement with empirical $T_{\rm eff}$
scale.  We advise that the large atomic line lists containing many 
theoretically predicted lines of Fe-group elements be used with caution.  

\paragraph{}

%  PHOENIX includes $b-f$ opacity from H, H$^{\rm -}$, \ion{He}{1} and \ion{}{2}, \ion{He}$^{\rm -}$, \ion{C}{1}, 
%C$^{\rm -}$, \ion{N}{1},
%\ion{Na}{1}, \ion{Mg}{1} and \ion{}{2}, \ion{Al}{1}, \ion{Si}{1} and \ion{}{2}, \ion{S}{1},\ion{Ca}{1} and \ion{}{2},
%\ion{Fe}{1}, \ion{O}{1}, H$_{\rm 2}$, H$_{\rm 2}^{\rm +}$, $f-f$ opacity from \ion{H}{1}, H$^{\rm -}$, H$_{\rm 2}^{\rm -}$,
% \ion{Mg}{1}, 
%and \ion{Si}{1}, Rayleigh scattering from \ion{H}{1}, \ion{He}{1}, and H$_{\rm 2}$, and Phillip & Phillip??? molbf?? (MgH b-f?.

PHOENIX includes continuous molecular opacity from  H$_{\rm 2}$, H$_{\rm 2}^{\rm +}$, 
H$_{\rm 2}^{\rm -}$ and MgH.  \citet{kurucz_dt87} found that including the photo-dissociation 
opacity of CH makes a detectable difference to the predicted flux of the Sun in a localized region
around 4000 \AA. Their calculations did not include the effect of line blanketing, so it is
difficult to judge how detectable the difference is.  Nevertheless, because molecules become
increasingly important as $T_{\rm eff}$ decreases it would be worthwhile investigating the
role of this opacity source in our models.  In any case, \citet{shorth09} found that the 
continuous opacity in the Sun and Arcturus in the 
$\lambda\lambda 3000$ to $4000$ region is dominated, variously, by
by H$^{\rm -}$ $b-f$, the combined $b-f$ opacity due to metals, the combined $f-f$ opacity of
 Mg and Si, and Thomson scattering, and the cross-sections for most of these
are kown accurately enough that their dominance is not in doubt.

\acknowledgments

CIS is grateful for NSERC Discovery Program grant 264515-07.  The calculations were
performed with the facilities of the Atlantic Computational Excellence Network (ACEnet).

\clearpage

%% the file, uncomment the \plotone, \plottwo, and
%% If you need a layout that cannot be achieved with \plotone or
%% \plottwo, you can invoke the graphicx package directly with the
%% \includegraphics command or use \plotfiddle. For more information,
%% please see the tutorial on "Using Electronic Art with AASTeX" in the
%% documentation section at the AASTeX Web site,
%% http://www.journals.uchicago.edu/AAS/AASTeX.
%%
%% The examples below also include sample markup for submission of
%% supplemental electronic materials. As always, be sure to check
%% the instructions to authors for the journal you are submitting to
%% for specific submissions guidelines as they vary from

%% This example uses \plotone to include an EPS file scaled to
%% 80% of its natural size with \epsscale. Its caption
%% has been written to indicate that additional figure parts will be
%% available in the electronic journal.

\clearpage

\clearpage

\clearpage

\clearpage

\clearpage

\clearpage

\clearpage

\clearpage

\clearpage

\clearpage

\clearpage

\clearpage

\clearpage

\clearpage

\clearpage

%% Tables should be submitted one per page, so put a \clearpage before

%% Two options are available to the author for producing tables:  the
%% deluxetable environment provided by the AASTeX package or the LaTeX
%% table environment.  Use of deluxetable is preferred.
%%

%% Tables may also be prepared as separate files. See the accompanying
%% sample file table.tex for an example of an external table file.
%% To include an external file in your main document, use the \input
%% command. Uncomment the line below to include table.tex in this
%% sample file. (Note that you will need to comment out the \documentclass,
%% \begin{document}, and \end{document} commands from table.tex if you want
%% to include it in this document.)


%% The following command ends your manuscript. LaTeX will ignore any text

\clearpage

\begin{deluxetable}{llrcll}
\tablecolumns{6}
\tablecaption{List of stars selected from B85 catalog for SED fitting.  Spectral types are from the sources cited in Section \ref{sobsseds}, $V$ values are from BSC5, and the [${{\rm A}\over{\rm H}}$] values are from \citet{cayrelsr01}. }
\tablehead{
\colhead{ }           & \colhead{Spectral} & \colhead{ }   & \colhead{Mean or median}            & \colhead{Num}                       & \colhead{Num} \\ 
\colhead{Object}      & \colhead{type}     & \colhead{$V$} & \colhead{[${{\rm A}\over{\rm H}}$]} & \colhead{[${{\rm A}\over{\rm H}}$]} & \colhead{spectra} 
} 
\startdata
HD50522   & G5 III-IV & 4.35 &  0.05  &  1  & 1\\
HD71369   & G5 III    & 3.36 & -0.05  &  2  & 2\\
%%%%%%%%%%%%HD29291   & G8IIIa   & 3.82 & -0.09  &  1  & 1\\
HD34559   & G8 III    & 4.94 & -0.10  &  1  & 1\\
HD62345   & G8 IIIa   & 3.57 &  0.00  &  1  & 3\\
HD100407  & G7 III    & 3.54 & -0.04  &  1  & 1\\
HD115659  & G8 IIIa  & 3.00 & -0.02  &  2  & 1\\
HD147675  & G8 K0III & 3.89 & -0.05  &  1  & 1\\
%%%%%%%%%%%%%HD148387  & G8-IIIab & 2.74 & -0.07  &  2  & 2\\
HD192947  & G8 IIIb   & 3.57 & -0.03  &  2  & 1\\
HD7106\tablenotemark{l} & K0.5 IIIb & 4.51 & -0.04  &  1  & 2\\
%%%%%%%%%%%%%HD8512    & K0III    & 3.60 & -0.05  &  2  & 2\\
HD61935\tablenotemark{h} & K0 III    & 3.93 & -0.10  &  2  & 2\\
HD62509\tablenotemark{h} & K0 IIIb   & 1.14 & -0.10  &  7  & 2\\
HD123139\tablenotemark{h} & K0 IIIb   & 2.06 &  0.04  &  2  & 1\\
HD125351\tablenotemark{l} & K0 III    & 4.81 &  -0.08  &  2  & 1\\
HD177241\tablenotemark{h} & G9 IIIb   & 3.77 &  0.03  &  2  & 1\\
HD188947\tablenotemark{h} & K0 III    & 3.89 &  0.05  &  2  & 1\\
HD205512\tablenotemark{l} & K0.5 III  & 4.90 &  0.05  &  4  & 2\\
HD216228\tablenotemark{h} & K0 III    & 3.52 &  0.03  &  3  & 1\\
HD219449\tablenotemark{l} & K0 III    & 4.21 &  -0.08 &  2  & 1\\
%%%%%%%%%%%%%HD47205   & K1III    & 3.95  & 0.04 & 3 & 1\\
HD62044   & K1 III    & 4.28 & -0.02 & 1 & 1\\
HD71878   & K1 III    & 3.77 & -0.01 & 1 & 1\\
%%%%%%%%%%%%%HD96833   & K1III    &  & -.18--.07 & 3 & 1\\
HD206952  & K1 III    & 4.56 & 0.04  & 1 & 1\\
HD156266  & K2 III    & 4.73 & -0.03  & 1 & 1\\
HD161096  & K2 III    & 2.77 & 0.06  & 2 & 2\\
HD49878   & K4 III  & 4.55 & 0.05      & 1 & 1\\
HD81817   & K3 IIIa & 4.29 & 0.09      & 1 & 1\\
%%%%%%%%%%%%%HD168454  & K3-IIIa &  & -0.01,-0.34,-0.32 & 3 & 1\\
HD211073  & K3 III  & 4.49 & -0.07 & 2 & 1\\
HD221673  & K4 IIIb & 4.98 & -0.03      & 1 & 1\\
\hline
HD115383  & G0 V    & 5.22 & 0.08   & 5 & 1\\
HD141004  & G0 V    & 4.43 & -0.02  & 6 & 1\\
HD20630   & G5 V    & 4.83 & 0.01   & 3 & 1\\
HD117176  & G4 V    & 4.98 & -.08   & 4 & 1\\
\hline
%%%%%%%%%%%%%HD5516    & G8IIIb  &  & -0.54     & 1 & 2\\
HD138905  & G8.5 III  & 3.91 & -0.41     & 2 & 1\\ 
HD222107  & G8 III-IV & 3.82 & -0.49     & 2 & 2\\
HD124897\tablenotemark{a} & K1.5 IIIFe-0.5 & -0.04 & -0.54 & 17 & 3\\ 
\enddata
\tablenotetext{l}{K0 III stars of ``low'' (l) UV $f_\lambda$ level; see text.}
\tablenotetext{h}{K0 III stars of ``high'' (h) UV $f_\lambda$ level; see text.}
\tablenotetext{a}{Arcturus, $\alpha$ Boo}
\label{tabb85}
\end{deluxetable}

%%%%%%%%%%%\clearpage
%%%%%%%%%%% This version: print RMS and deviations for total range, and blue & red sub-ranges of best *global* fit
%%%%%%%%%%%\begin{deluxetable}{lrrrrrrr}
%%%%%%%%%%%\tablecaption{Closest match models to mean sample spectra and goodness of fit statistics. }
%%%%%%%%%%%\tablehead{\colhead{Sample spectral type} & \colhead{$T_{\rm eff}$} & \colhead{$\log g$} & \colhead{[${{\rm A}\over{\rm H}}$]} & \colhead{$\sigma$} & \colhead{$\sigma_{\rm blue}$} & \colhead{$\sigma_{\rm red}$} & \colhead{$\Delta_{\rm blue}$} & \colhead{$\Delta_{\rm red}$} } 
%%%%%%%%%%%\startdata
%%%%%%%%%%%G5 III & 5313 & 2.5 & 0.0 & 0.0469 & 0.0579 & 0.0390 & 0.0189 & -0.0320 \\
%%%%%%%%%%%G8 III & 5125 & 2.5 & 0.0 & 0.0379 & 0.0604 & 0.0119 & 0.0271 & -0.0071 \\   
%%%%%%%%%%%K0 III & 4875 & 2.5 & 0.0 & 0.0529 & 0.0836 & 0.0187 &-0.0187 & -0.0138 \\  
%%%%%%%%%%%K0 III^{\rm h} & 4938 & 2.5 & 0.0 & 0.0471 & 0.0735 & 0.0190 & -0.0122 & -0.0148 \\
%%%%%%%%%%%K0 III^{\rm l} & 4875 & 2.0 & 0.0 & 0.0548 & 0.0846 & 0.0243 & -0.0120 & -0.0141\\
%%%%%%%%%%%K1 III & 4687 & 2.5 & 0.0 & 0.07914 & 0.1252 & 0.0278 & -0.0227 & -0.0201 \\     
%%%%%%%%%%%K2 III & 4625 & 2.0 & 0.0 & 0.0580 & 0.0878 & 0.0288 & -0.0150 & -0.0201 \\   
%%%%%%%%%%%K4 III  
%%%%%%%%%%%\\
%%%%%%%%%%%G0 V    
%%%%%%%%%%%G5 V    
%%%%%%%%%%%\\
%%%%%%%%%%%G8 III  
%%%%%%%%%%%\enddata
%%%%%%%%%%%\label{tabstats}
%%%%%%%%%%%\end{deluxetable} 

\clearpage

%% This version: print RMS best fit stellar parameters for total range,and blue & red sub-ranges 
%
%% BIG line list
\begin{deluxetable}{lrrrrr}
\tablecolumns{6}
\tablecaption{Series 1 models: Closest match models to mean sample spectra and goodness of fit statistics. }
\tablehead{
\colhead{}              & \multicolumn{2}{c}{Total SED}       &                     \colhead{Blue}                     & \colhead{Red}                      & \colhead{} \\ 
\colhead{Spectral type} & \colhead{$T_{\rm eff}$ ($\sigma$)} & \colhead{$\log g$} & \colhead{$T_{\rm eff}$ ($\sigma$)} & \colhead{$T_{\rm eff}$ ($\sigma$)} & \colhead{[${{\rm A}\over{\rm H}}$]} 
} 
\startdata
%%G5 III & 5312 (0.047) & 2.5 & 5312 (0.058) & 5188 (0.024) & 0.0  \\
G5 III & \nodata\tablenotemark{a} & 2.5 & 5312 (0.058) & \nodata\tablenotemark{a} & 0.0  \\
G8 III & 5125 (0.039) & 2.5 & 5125 (0.058) & 5062 (0.013) & 0.0 \\   
K0 III & 4875 (0.049) & 2.0 & 4875 (0.077) & 4812 (0.012) & 0.0 \\  
K0 III\tablenotemark{h} & 4938 (0.046) & 2.0 & 4938 (0.069) & 4875 (0.016) & 0.0 \\
K0 III\tablenotemark{l} & 4812 (0.059) & 2.0 & 4812 (0.093) & 4750 (0.016) & 0.0\\
K1 III & 4688 (0.079) & 2.5 & 4688 (0.125) & 4625 (0.014) & 0.0 \\    
K2 III & 4562 (0.062) & 2.0 & 4562 (0.098) & 4500 (0.017) & 0.0  \\   
K3-4 III & 4250 (0.109) & 2.0 & 4250 (0.169) & 4188 (0.041) & 0.0 \\   
\hline
G0 V   & 5938 (0.039) & 4.5 & 5938 (0.059) & 5812 (0.014) & 0.0 \\   
G4-5 V & 5688 (0.049) & 4.5 & 5688 (0.075) & 5625 (0.021) & 0.0 \\
\hline
G8 III                   & 4688 (0.063) & 2.5 & 4688 (0.094) & 4562 (0.013) & -0.5 \\  
K1.5 III\tablenotemark{b} & 4250 (0.095) & 2.0 & 4250 (0.147) & 4312. (0.026) & -0.5 \\
\enddata
\tablenotetext{a}{See text.}
\tablenotetext{b}{Arcturus, $\alpha$ Boo}
\tablenotetext{h}{K0 III sub-sample of ``high'' UV flux (see text).}
\tablenotetext{l}{K0 III sub-sample of ``low'' UV flux (see text).}
\label{tabstats1}
\end{deluxetable} 

\clearpage

%% SMALL line list
\begin{deluxetable}{lrrrrr}
\tablecolumns{6}
\tablecaption{Series 2 models: Same as Table \ref{tabstats1}. }
\tablehead{
\colhead{}              & \multicolumn{2}{c}{Total SED}                           & \colhead{Blue}                     & \colhead{Red}                      & \colhead{} \\ 
\colhead{Spectral type} & \colhead{$T_{\rm eff}$ ($\sigma$)} & \colhead{$\log g$} & \colhead{$T_{\rm eff}$ ($\sigma$)} & \colhead{$T_{\rm eff}$ ($\sigma$)} & \colhead{[${{\rm A}\over{\rm H}}$]} 
} 
\startdata
%%%%%%%%%%%%G5 III & 5312 (0.049) & 2.5 & 5312 (0.065) & 5188 (0.022) & 0.0 \\
G5 III & \nodata\tablenotemark{a} & 2.5 & 5312 (0.065) & \nodata\tablenotemark{a} & 0.0 \\
G8 III & 5125 (0.037) & 2.0 & 5125 (0.057) & 5062 (0.012) & 0.0 \\
K0 III & 4812 (0.051) & 2.0 & 4812 (0.082) & 4812 (0.012) & 0.0 \\
K0 III\tablenotemark{h} & 4875 (0.046) & 2.0 & 4875 (0.074) & 4875 (0.014) & 0.0 \\ 
K0 III\tablenotemark{l} & 4750 (0.057) & 2.0 & 4750 (0.090) & 4750 (0.018) & 0.0 \\ 
K1 III & 4625 (0.074) & 2.5 & 4625 (0.121) & 4625 (0.015) & 0.0 \\ 
K2 III & 4500 (0.059) & 2.0 & 4500 (0.095) & 4500 (0.018) & 0.0 \\ 
K3-4 III & 4250 (0.101) & 1.5 & 4250 (0.155) & 4188 (0.036) & 0.0 \\ 
\hline
G0 V   & 5938 (0.048) & 4.5 & 5938 (0.076) & 5875 (0.012) & 0.0 \\ 
G4-5 V & 5625 (0.060) & 4.5 & 5625 (0.094) & 5688 (0.022) & 0.0 \\  
\hline
G8 III                   & 4688 (0.058) & 2.5 & 4688 (0.086) & 4562 (0.013) & -0.5 \\ 
K1.5 III\tablenotemark{b} & 4250 (0.096) & 2.0 & 4250 (0.149) & 4312 (0.027) & -0.5 \\
\enddata
\tablenotetext{a}{See text.}
\tablenotetext{b}{Arcturus, $\alpha$ Boo}
\tablenotetext{h}{K0 III sub-sample of ``high'' UV flux (see text).}
\tablenotetext{l}{K0 III sub-sample of ``low'' UV flux (see text).}
\label{tabstats2}
\end{deluxetable}

%\clearpage
%%Compare to empirical calibration of Ramirez & Melendez (Alonso lineage) (and to theoretical w. NextGen of Bertone, Buzzoni, et al. 
%
%% BIG line list
\begin{deluxetable}{lrrrrrrrr}
\tablecolumns{9}
\tablecaption{Comparison with empirical $T_{\rm eff}$ calibrations of RM05 and BBCR04.}
\tablehead{
\colhead{ }             & \colhead{ }   & \multicolumn{2}{c}{Series 1}     & \multicolumn{2}{c}{Series 2}     & \colhead{ }    & \colhead{ } \\ 
\colhead{Spectral type} & \colhead{B-V} ($\sigma$) & \colhead{Blue } & \colhead{Red } & \colhead{Blue } & \colhead{Red } & \colhead{RM05} & \colhead{BBCR04} & \colhead{B10} } 
\startdata
G5 III & 0.882 (0.019) & 5312 & \nodata & 5312 & \nodata & 5137 & 5410 & \nodata \\
G8 III & 0.930 (0.004) & 5125 & 5062 & 5125 & 5062 & 4964 & 5123\tablenotemark{a}& \nodata  \\   
K0 III & 1.043 (0.002) & 4875 & 4812 & 4812 & 4812 & 4721 & 4870 & \nodata\\  
K0 III\tablenotemark{h} & 1.018 (0.011) & 4938 & 4875 & 4875 & 4875 & 4781 & 4870 & \nodata \\
K0 III\tablenotemark{l} & 1.080 (0.000) & 4812 & 4750 & 4750 & 4750 & 4650 & 4870& \nodata\\
K1 III & 1.115 (0.003) & 4688 & 4625 & 4625 & 4625 & 4592 & 4710\tablenotemark{a} & 4737 \\     
K2 III & 1.160 (0.006) & 4562 & 4500 & 4500 & 4500 & 4531 & 4550  & 4562 \\   
K3-4 III & 1.408 (0.014) & 4250 & 4188 & 4250 & 4188 & 4118 & 4243\tablenotemark{a} & 4134 \\   
\hline
G0 V & 0.595 (0.004) & 5983 & 5812 & 5938 & 5875 & 5864 & 6140 & \nodata \\   
G4-5 V & 0.695 (0.010) & 5688 & 5625 & 5625 & 5688 & 5519 & 5670& \nodata\\
\hline
G8 III-0.5 & 1.010 (0.000) & 4688 & 4562 & 4688 & 4562 & 4684 & \nodata & \nodata\\  
K1.5 III-0.5$^{\rm b}$ & 1.211 (0.009) & 4250 & 4312 & 4250 & 4312 & 4332 & \nodata & 4386\\     
%%%%%%%%%%%%G5 III & 0.843,0.854,0.849 & 5313 & 5188 &  5137  \\
%%%%%%%%%%%%G8 III & 0.925,0.919,0.922 & 5125 & 5125 &  4973 \\   
%%%%%%%%%%%%K0 III & 1.040,1.040,1.040 & 4875 & 4813 &  4735 &\\  
%%%%%%%%%%%%K0 III^{\rm h} & 1.008,1.007,1.007 & 4938 & 4875 & 4798  \\
%%%%%%%%%%%%K0 III^{\rm l} & 1.066,1.067,1.067 & 4813 & 4813 & 4684 \\
%%%%%%%%%%%%K1 III & 1.117,1.119,1.118 & 4688 & 4625 & 4592 \\     
%%%%%%%%%%%%K2 III & 1.145,1.138,1.142 & 4625 & 4563 & 4550  \\   
%%%%%%%%%%%%K4 III & 1.362,1.364,1.363 & 4313 & 4250 & 4188 \\   
%%%%%%%%%%%%\\
%%%%%%%%%%%%%G0 V & 0.595,0.595,0.595 & 5938 & 5813 & 5864  \\   
%%%%%%%%%%%%G5 V & 0.695,0.698,0.697 & 5688 & 5625 & 5519 \\
%%%%%%%%%%%\\
%%%%%%%%%%%%G8 III & 1.010,0.992,1.001 & 4688 & 4563 & 4684 \\  
\enddata
\tablenotetext{a}{$T_{\rm eff}$ values found from linear interpolation in Table 1 of BBCR04.}
\tablenotetext{b}{Arcturus, $\alpha$ Boo}
\tablenotetext{h}{K0 III sub-sample of ``high'' UV flux (see text).}
\tablenotetext{l}{K0 III sub-sample of ``low'' UV flux (see text).}
\label{tabcomp1}
\end{deluxetable} 

%%%%%%%%%%\clearpage
%%%%%%%%%%
%%%%%%%%%%%% SMALL line list
%%%%%%%%%%%\begin{deluxetable}{lrrrrr}
%%%%%%%%%%%\tablecaption{Series 1 models: Comparison with empirical $T_{\rm eff}$ calibration of \citet{ramirezm05} and \citet{Bertoneb04}.}
%%%%%%%%%%%\tablehead{\colhead{Spectral type} & \colhead{B-V} & \colhead{Blue range} & \colhead{Red range} & \colhead{RM05} & \colhead{BBCR04} } 
%%%%%%%%%%%\startdata
%%%%%%%%%%%G5 III & 5313 & 5188 &  \\
%%%%%%%%%%%G8 III & 5125 & 5125 &  \\
%%%%%%%%%%%K0 III & 4875 & 4813 &  \\
%%%%%%%%%%%K0 III^{\rm h} & 4938 & 4938 &  \\ 
%%%%%%%%%%%K0 III^{\rm l} & 4813 & 4813 &  \\ 
%%%%%%%%%%%K1 III & 4625 & 4625 &  \\ 
%%%%%%%%%%%K2 III & 4563 & 4563 &  \\ 
%%%%%%%%%%%K3-4 III & 4250 & 4250 &  \\ 
%%%%%%%%%%%\\
%%%%%%%%%%%G0 V & 5938 & 5875 &  \\ 
%%%%%%%%%%%G4-5 V & 5625 & 5688 &  \\  
%%%%%%%%%%%\\
%%%%%%%%%%%G8 III & 4688 & 4563 &  \\ 
%%%%%%%%%%%\enddata
%%%%%%%%%%%\label{tabcomp2}
%%%%%%%%%%%\end{deluxetable} 

\end{document}